\documentclass[12pt]{article}
\usepackage[utf8]{inputenc}
\usepackage[english]{babel}
\usepackage{graphicx}     % images (PNG/JPG/PDF)
\usepackage{float}        % for [H]
\usepackage[a4paper,left=1in,right=1in,top=1in,bottom=1.2in]{geometry}
\usepackage{amsmath, amssymb, amsthm, mathtools}
\usepackage{array}
\usepackage{booktabs}
\usepackage{tabularx}
\usepackage{threeparttable}
\usepackage{multirow}
\usepackage{makecell}
\usepackage{enumerate}
\usepackage{siunitx}
\usepackage{xcolor}
\usepackage{tikz}

\usepackage{subcaption}
\usepackage{hyperref}

\title{Statistical Properties of the Rooted-Tree Encoding of $\mathbb{N}$}

\author{
  Pierluigi Contucci\textsuperscript{1}\thanks{Email: pierluigi.contucci@unibo.it}
  \and
  Claudio Giberti\textsuperscript{2}\thanks{Email: claudio.giberti@unimore.it}
  \and
  Godwin Osabutey\textsuperscript{3}\thanks{Corresponding author. Email: gosabutey@unimore.it}
  \and
  Cecilia Vernia\textsuperscript{3}\thanks{Email: cecilia.vernia@unimore.it}
}

\date{
\textsuperscript{1}\textit{Department of Mathematics, University of Bologna,  
Via Zamboni 33, 40126 Bologna, Italy}\\[5pt]
\textsuperscript{2}\textit{Department of Sciences and Methods for Engineering,  
University of Modena and Reggio Emilia,  
Via G. Amendola 2, 42122 Reggio Emilia, Italy}\\[5pt]
\textsuperscript{3}\textit{Department of Physics, Informatics and Mathematics,  University of Modena and Reggio Emilia,  
Via G. Campi 213/b, 41125 Modena, Italy}\\[12pt]
\today
}

\newcommand{\Nat}{\(\mathbb{N}\)atural Text }
\newcommand{\Natv}{\(\mathbb{N}\)atural Text}

\newcommand{\Ntv}{\(\mathbb{N}\mathcal{T}\)}
\newcommand{\Nad}{\(\mathbb{N}\)atural Dictionary }

\newcommand{\Nd}{\(\mathbb{N}\mathcal{D}\)}
\newcommand{\invisible}[1]{}

\begin{document}

%\date{\today}
\maketitle

\begin{abstract}
We prime-encode the natural numbers via recursive factorisation, iterated to the exponents, generating a corpus of planar rooted trees  equivalently represented as Dyck words. This  forms a deterministic text endowed with internal rules. Statistical analysis of the corpus reveals that the dictionary and the entropy grow sublinearly, compression shows non–monotonic trend, and the rank--frequency curves assume a stable parabolic form deviating from Zipf's law. Correlation analysis using mean-squared displacement reveals a transition from normal diffusion to superdiffusion in the associated walk. These findings characterise the tree-encoded sequence as a statistically structured text with long–range correlations grounded in its generative arithmetic law, providing an empirical basis for subsequent theoretical and learnability investigations.
\end{abstract}


\section{Introduction}

The sequence of natural numbers, when expressed through their iterated prime
factorisation, gives rise to an ordered chain of rooted trees, a purely arithmetic
structure that can be read as a symbolic text. In this representation, each number is
translated into a rooted tree whose branches encode the Euclidean recursive
decomposition \cite{Childress_2021,Stanley2011}, and the resulting corpus, once the prime
labels are discarded, becomes a deterministic language written in Dyck words.

In this work, we treat that sequence as an empirical object. Rather than imposing a
generative model, a stochastic hypothesis, or any notion of randomness, we record
its observables directly: the growth of the dictionary, the rank–frequency
distribution of tree types, the symmetry properties of the text, its entropy and
compressibility, and the correlation structure extracted via associated walks. The
approach is descriptive in the strict sense: the analysis is restricted to what is
measured, without interpretation beyond the arithmetic process that generates the
data.

The findings are consistent across scales. The dictionary of distinct Dyck words
grows sublinearly, indicating a systematic reuse of structures and suggesting an
implicit combinatorial grammar. The entropy increases with corpus size, yet remains
well below the non–informative maximum, and the compression ratio exhibits a
non–monotonic behaviour indicative of structural organization. The rank–frequency
curve stabilizes into a parabolic form in log–log scale, reminiscent of hierarchical
self–similar regimes found in other correlated corpora
\cite{Sornette1997,Montemurro2001}. Finally, the
analysis of walks associated with specific Dyck words reveals a two–regime
behaviour in the mean–square displacement, from normal diffusion at short time scales to
superdiffusion or quasi–ballistic at longer ranges, suggesting the presence of
long–range deterministic correlations.

Together, these results provide an empirical characterization of the arithmetic text
as a structured, statistically organized language emerging from a fully
deterministic process. The present study establishes the descriptive foundation for
subsequent theoretical work and for investigations of learnability, in particular by
computational models trained on this corpus. 

This paper is structured as follows. Section~\ref{sec: tree encoding} introduces the tree encoding and the associated Dyck word representation. Section~\ref{sec:text and dic} reports the main statistical observations: dictionary growth and lexical reuse, directionality, entropy and compressibility, rank–frequency structure, and dynamical features captured through the mean–square displacement and cross–correlation of Dyck words.  Section \ref{sec:conclusion} summarizes the conclusions and suggests future research avenues that emerge naturally from the results.

\section{Tree Encoding and Dictionary Analysis}\label{sec: tree encoding}
The database studied in this paper is obtained by prime factorisation of natural numbers, according to the Fundamental Theorem of Arithmetic. More precisely, every 
\begin{equation}\label{eq:natural}
n \in \mathbb{N} = \{1,2,3,\dots\}, \quad n>1,
\end{equation}
can be uniquely written as  
\begin{equation}\label{eq:factorisation}
n = p_1^{n_1}  p_2^{n_2} \dots p_k^{n_k},
\end{equation}
where $p_j \in \mathbb{P}=\{2,3,5,\dots\}$ are prime numbers and $n_j \in \mathbb{N}$, with  
\begin{equation}\label{eq:ordering}
p_1 < p_2 < \dots < p_k \, .
\end{equation}
By convention, 
\begin{equation}\label{eq:unit}
n=1
\end{equation}
corresponds to the empty product of primes.

In order to obtain a \textit{pure–prime} representation of a natural number, the factorisation process may be iterated by decomposing the exponents in \eqref{eq:factorisation} into their own prime factorisations, and then proceed recursively \cite{Iudelevich_2022}. In this way, every natural number is represented through primes only, both at the base and in the exponents. 

Then, for instance,
\begin{equation}\label{eq:small_examples}
12 = 2^2 \cdot 3, 
\qquad 
72 = 2^3 \cdot 3^2, 
\qquad 
320 = 2^6 \cdot 5 = 2^{2 \cdot 3} \cdot 5 ,
\end{equation}
or, in a more elaborate case,
\begin{equation}\label{eq:big_example}
3099363912 = 2^3 \cdot 3^{18} 
           = 2^3 \cdot 3^{2 \cdot 3^2}.
\end{equation}

In this way, we can \textit{prime-encode} every natural number, expressing it entirely in terms of primes at all levels, using only products and exponentiation.
\begin{figure}[H]
\centering
\begin{tikzpicture}[
  hollow node/.style={circle,draw,inner sep=1.5pt},
  solid node/.style={circle,draw,inner sep=1.5pt,fill=black}
]

\node(0)[solid node]{}
  % Left branch: 2 -> 3
  child[grow=120]{node[solid node]{}
    child[grow=120]{node[solid node]{} edge from parent node[xshift=-10]{$3$}}
    edge from parent node[xshift=-10]{$2$}
  }
  % Right branch: 3 -> bifurcates
  child[grow=60]{node[solid node]{}
    % Left child: 2
    child[grow=120]{node[solid node]{} edge from parent node[xshift=-10]{$2$}}
    % Right child: 3 -> 2
    child[grow=60]{node[solid node]{}
      child[grow=60]{node[solid node]{} edge from parent node[xshift=10]{$2$}}
      edge from parent node[xshift=10]{$3$}
    }
    edge from parent node[xshift=10]{$3$}
  };
\end{tikzpicture}
\vskip 0.60cm
\caption{Decorated tree representation of the integer $3099363912 = 2^3 \cdot 3^{18} = 2^3 \cdot 3^{2 \cdot 3^2}$}\label{fig:bigtree}
\end{figure}
It is a simple, yet remarkable observation that every natural number corresponds uniquely to a prime-decorated planar rooted tree (see Figure \ref{fig:bigtree}). This correspondence can be regarded as a tree-theoretic refinement of the Fundamental Theorem of Arithmetic. 

In this framework, primes correspond to the simplest planar rooted tree consisting of a single edge, and the study of the distribution of prime numbers becomes the study of the occurrence of such \textit{undecorated} trees within the natural sequence. We then consider the sequence of all undecorated trees associated with successive natural numbers. The first ten examples are shown below:
\begin{figure}[H]
    \centering
    \includegraphics[width=0.95\linewidth]{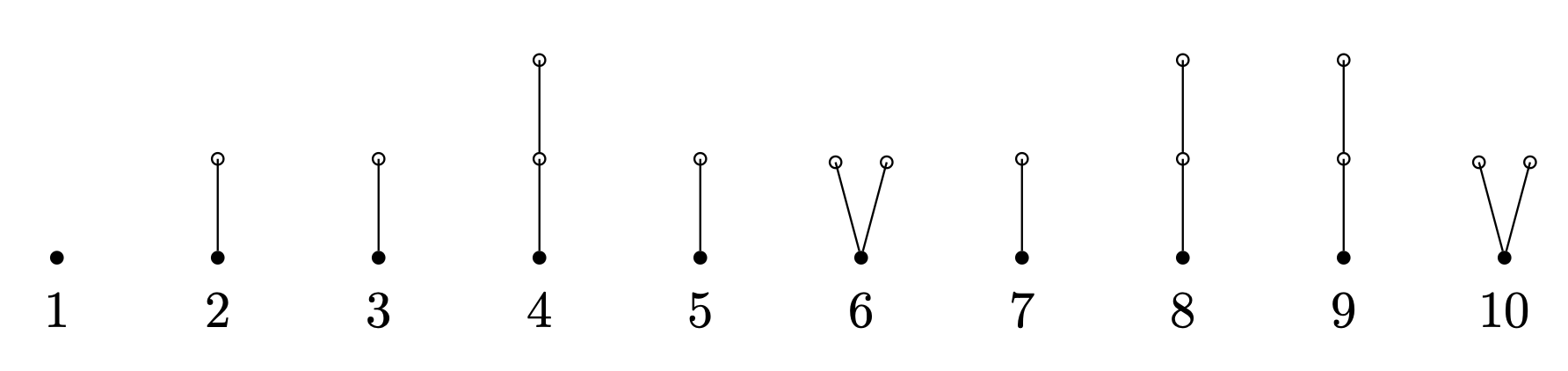}
\end{figure}
Apart from the empty tree, only three distinct types appear among the first ten natural numbers: those corresponding to primes $p$, to products of two primes $p\cdot q$, and to natural numbers of the form $p^q$. 

For analytical purposes it is convenient to replace trees by their \textit{Dyck} word encodings, namely balanced binary sequences of \texttt{1}'s and \texttt{0}'s. 
The one-to-one map associating a planar rooted tree to its Dyck sequence is obtained by circumnavigating the tree clockwise writing a \texttt{1} whenever the path moves up and a \texttt{0} whenever it moves down. Here is an explicit example:
\begin{figure}[H]
    \centering \includegraphics[width=0.35\linewidth, height=4cm]{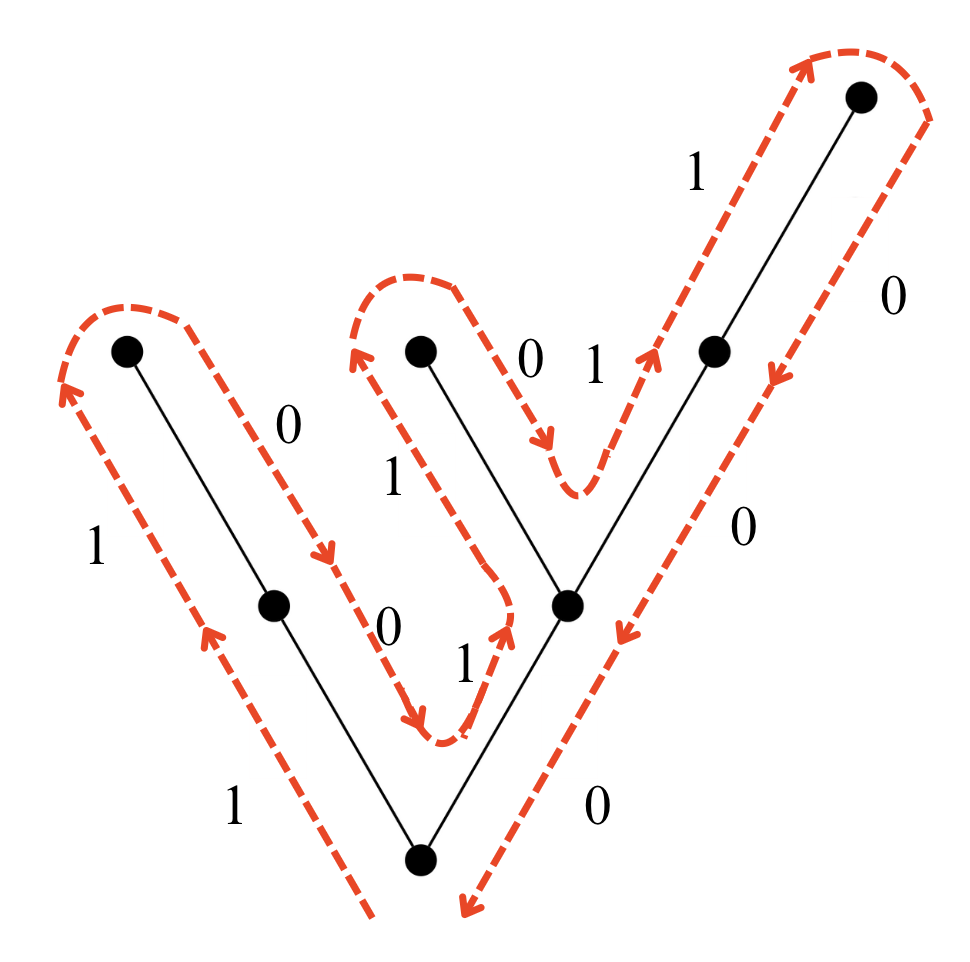}
\caption{A tree and the corresponding Dyke word \texttt{110011011000}.}
\label{fig:dycktree}
\end{figure}
This generates a natural sequence of Dyck words, and the first ten of them are 
\begin{equation}\label{eq:nattext-1-10}
\texttt{$\emptyset$, 10, 10, 1100, 10, 1010, 10, 1100, 1100, 1010}
\end{equation}

In \cite{ Conti_Contucci_Iudelevich_2024, Conti_Contucci_2025} it was observed that the infinite text of Dyck words exhibits a number of striking structural properties. 
Since there are infinitely many primes, each Dyck word appears infinitely many times throughout the sequence. 
Nevertheless, certain short \textit{phrases} (finite sequences of Dyck words) occur only once, such as \texttt{10 10} or $\texttt{1100 10}$, while others recur extremely often. 
For instance, the infinite occurrence of $\texttt{10 1100}$ is equivalent to the still open Mersenne conjecture, the infinity of $\texttt{1010 10}$ relates to the Sophie Germain conjecture, see, e.g.\cite{Conti_Contucci_2025}, and $\texttt{1100 1100}$ is known to occur once, a fact established in 2004 with the proof of Catalan's conjecture \cite{Mihailescu2004}. 

The text is moreover oriented: if one takes a valid phrase and flips its order, the resulting string is often invalid, i.e. it never occurs. For example, $\texttt{111000 10}$ occurs multiple times, whereas $\texttt{10 111000}$ never appears.  
Overall, the text displays a complex structure and, in principle, an infinite number of rules that depend only on the underlying tree structure and not on its decoration. 
This naturally raises the question whether the sequence can be regarded as a \textit{text} and, if so, whether such a text can be \textit{learned} with the help of transformers used in Large Language Models of Artificial Intelligence. 

We define $\mathbb{N}\mathcal{T}$ ($\mathbb{N}$atural Text) the sequence of Dyck words and 
$\mathbb{N}\mathcal{D}$ ($\mathbb{N}$atural Dictionary) the ordered sequence of all the different Dyck words.
This paper is devoted to the statistical properties of the two sets up to a database of factorised natural numbers of $6.5$ billions.

\section{Text and Dictionary Statistical Analysis}\label{sec:text and dic}
Given the \Nat \Ntv, we denote ${\mathbb{N}\mathcal{T}}_{\ell}^{n}$ the part of the text found from the position $\ell$ to $n$, and ${\mathbb{N}\mathcal{D}}_{\ell}^{n}$ the $\mathbb{N}$atural Dictionary found in the same part of the \Natv. Then, for instance, ${\mathbb{N}\mathcal{T}}_{1}^{10}$ is the set of Dyck words given in \eqref{eq:nattext-1-10} and 
${\mathbb{N}\mathcal{D}}_{1}^{10}=(\texttt{10, 1100, 1010})$ (the Dyck word $\emptyset$ is not included in the \Nad since it appears only once in the sequence). Obviously,  the \Nad size, i.e. its cardinality, satisfies $|{\mathbb{N}\mathcal{D}}_{1}^{n} | \le | {\mathbb{N}\mathcal{T}}_{1}^{n} |=n$. In the sequel we denote by $d_n$ the length of ${\mathbb{N}\mathcal{D}}_{1}^{n}$.
We are now going to analyse ${\mathbb{N}\mathcal{T}}_1^{6.5\cdot 10^9}$, i.e. the dataset of Dyck words that has been obtained by the complete prime-encoding of the natural numbers up to $\overline{N}=6.5 \cdot 10^9$. 

\subsection{Dictionary Density}\label{sez:DD}
\begin{figure}[h]
    \centering
\includegraphics[width=10cm]{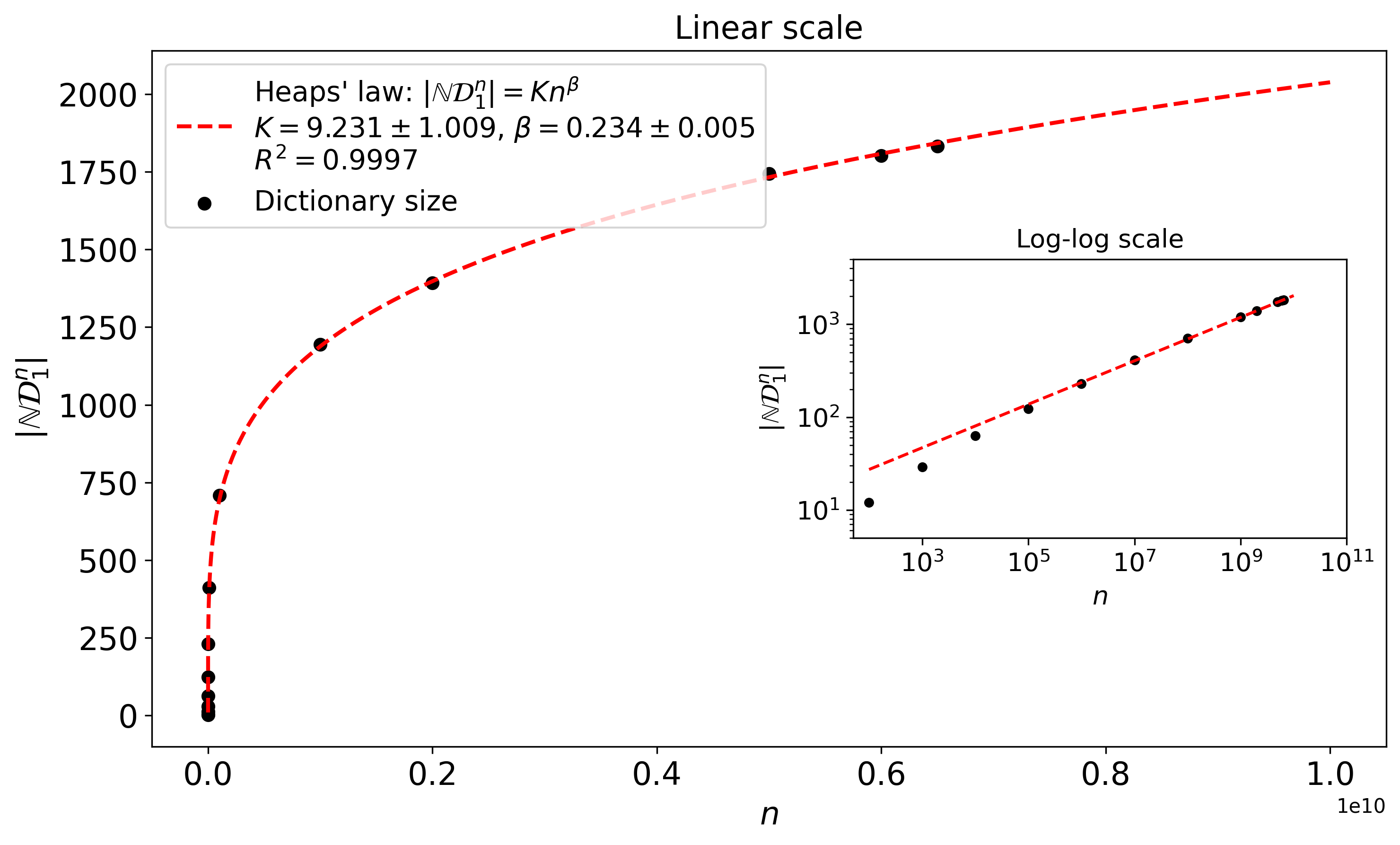
}
\caption{The number of different Dyck words $d_n=|{\mathbb{N}\mathcal{D}}_{1}^{n} |$ in the $\mathbb{N}$atural Text from position 1 to $n$ is represented versus $n$ (black dots). The dashed line represents the power-law fit to the data. In the inset the same data are represented in a log-log plot.}
\label{heaps}
\end{figure}

Our analysis of ${\mathbb{N}\mathcal{T}}_1^{6.5\cdot 10^9}$ starts with the computation of the dictionary density, i.e. $\delta_n=d_n/n$, as $n$ increases. 
In Table \ref{tab:dictionary size} we report the length of the \Nad for some values of $n$, while in Figure \ref{heaps} $|{\mathbb{N}\mathcal{D}}_{1}^{n} |$ is represented versus $n$. The power-law fit in the same figure highlights a sub-linear behaviour (which in linguistics is a well known phenomenon called Heaps' law).  More precisely, within the available data window, i.e. ${\mathbb{N}\mathcal{T}}_1^{6.5\cdot 10^9}$, the best power-law fit for the dictionary amplitude is given by $d_n \sim Kn^{\beta}$, with $\beta=0.234\pm 0.005$, $K=9.231\pm 1.009$ and a coefficient of determination $R^2=0.9997$. Thus, while we do not assert that the density exhibits an asymptotic power-law behaviour, we may conclude that, within the set of the first $6.5 \cdot 10^9$ $\mathbb{N}$atural Text elements, the dictionary density is approximately described by $\delta_n \sim n^{-0.7656}$ for $n\le 6.5 \cdot 10^9$.  

\begin{table}[H]
\centering
\caption{$\mathbb{N}$atural Text length $n$, the associated dictionary size $d_n=|\mathbb{N}\mathcal{D}_{1}^{n}|$ and its density $\delta_n$.}
\label{tab:dictionary size}
\begin{tabular}{
    S[table-format=1.0]   % prima colonna: scritta manualmente come 10^{n}
    S[table-format=4.0]
    S[scientific-notation = true, exponent-product = \cdot, table-format=1.2e-1]
}
\toprule
{$n$} & {$d_n$} & {$\delta_n=d_n/n$} \\
\midrule
1 & 1 & 1e0 \\
{$10^{1}$} & 3 & 3.0e-1 \\
{$10^{2}$} & 12 & 1.2e-1 \\
{$10^{3}$} & 29 & 2.9e-2 \\
{$10^{4}$} & 63 & 6.3e-3 \\
{$10^{5}$} & 123 & 1.23e-3 \\
{$10^{6}$} & 230 & 2.3e-4 \\
{$10^{7}$} & 412 & 4.12e-5 \\
{$10^{8}$} & 708 & 7.08e-6 \\
{$10^{9}$} & 1195 & 1.195e-6 \\
\bottomrule
\end{tabular}
\end{table}

\subsection{Text Orientation}
As already highlighted in the introduction, the $\mathbb{N}$atural Text presents a specific orientation, as shown by some examples. In this subsection, our aim is to provide statistical quantitative evidence of that fact by showing how it manifests itself across multiple scales in the \Natv.
To do this, we compute the proportion of ordered sequences of consecutive Dyck words that are present in ${\mathbb{N}\mathcal{T}}_{1}^{n}$ together with their mirror-symmetric version. 

To clarify the definition, let us first illustrate the core idea with an example. Consider the following sequence:
\begin{equation}
{\mathbb{N}\mathcal{T}}_{2}^{10} = (\texttt{10, 10, 1100, 10, 1010, 10, 1100, 1100, 1010}).
\end{equation}
For such a sequence, we list all distinct contiguous subsequences $\tau$ (tuples) of a fixed length $k$ and check whether their mirror-reversed versions $\tau^{\rm rev}$ also appear in the same window ${\mathbb{N}\mathcal{T}}_{2}^{10}$ of the \Natv. For instance, considering the distinct 4-tuples in $\mathbb{N}\mathcal{T}_{2}^{10}$, one has that
\begin{equation}
\tau=(\texttt{10, 10, 1100, 10})\mbox{\, is a sub-sequence of } \mathbb{N}\mathcal{T}_{2}^{10}, 
\end{equation}
but its mirror-reversed version \quad 
\begin{equation}
\tau^{\rm rev}=(\texttt{10, 1100, 10, 10}) \mbox{\, is not a sub-sequence of\, }\mathbb{N}\mathcal{T}_{2}^{10}.
\end{equation}
Enumerating all distinct contiguous 4-tuples of Dyck words in $\mathbb{N}\mathcal{T}_{2}^{10}$, we find 6 of them in total, of which 2 have their mirror-reversed versions also present. The fraction $\frac{2}{6}$ quantifies the \emph{Statistical Symmetry} of the 4-tuples in the sequence ${\mathbb{N}\mathcal{T}}_{2}^{10}$. 

%Hence, we define the fraction $\frac{2}{6}$ as the \emph{Statistical Asymmetry} of the 4-tuples in the sequence ${\mathbb{N}\mathcal{T}}_{2}^{10}$ .

\par
We now generalize this concept. Let us denote by $x_i$ the Dyck words appearing in the \Natv. Then writing
\begin{equation}\label{eq:nattextx}
{\mathbb{N}\mathcal{T}}_{1}^{n}=(x_1,\ldots,x_n)     
\end{equation}
where $x_i\in
{\mathbb{N}\mathcal{D}}_{1}^{n}$, we can introduce, for $1\le k \le n$, the set formed by the {\em different} contiguous subsequences of length $k$ appearing in
${\mathbb{N}\mathcal{T}}_{1}^{n}$, that is:
\begin{multline}
\Pi_k({\mathbb{N}\mathcal{T}}_{1}^{n})=\{ (z_1,\ldots, z_k) \in ({\mathbb{N}\mathcal{D}}_{1}^{n})^k\, |\,  (z_1,\ldots, z_k)= (x_i,\ldots x_{i+k-1}) |\\
\mbox{ for some } 1 \le i \le n-k+1\}.
\end{multline}   
Then, for each observed $k$-tuple $\tau=(z_1,\ldots, z_k)\in \Pi_k({\mathbb{N}\mathcal{T}}_{1}^{n})$ we denote its reverse by
\begin{equation}
    \tau^{\rm rev}=(z_k,\ldots, z_1),
\end{equation}
and define the {\em Statistical Symmetry} of order $k$ at length $n$ as the fraction of observed $k$-tuples in ${\mathbb{N}\mathcal{T}}_{1}^{n}$ whose reverse also appears in the sequence:
\begin{equation}\label{eq:defSym}
    {\rm SS}(k,n)= \frac{1}{|\Pi_k({\mathbb{N}\mathcal{T}}_{1}^{n})|} \sum_{\tau\in \Pi_k({\mathbb{N}\mathcal{T}}_{1}^{n})} \mathbf{1}\! \Big(\tau^{\mathrm{rev}} \in \Pi_k({\mathbb{N}\mathcal{T}}_{1}^{n})\Big),\quad 1\le k \le n,
\end{equation}
where $\mathbf{1}(X)$ is the indicator of the event $X$.
In a strictly oriented text, i.e. a text  where no $k$-tuple has a reverse, this quantity is obviously zero, while in the opposite case in which each Dyck word appears in the text with its reverse, the Statistical Symmetry is one. In all other cases the quantity is between these two values, $0$ and $1$.\par
\begin{figure}[H]
    \centering
\includegraphics[width=11cm]{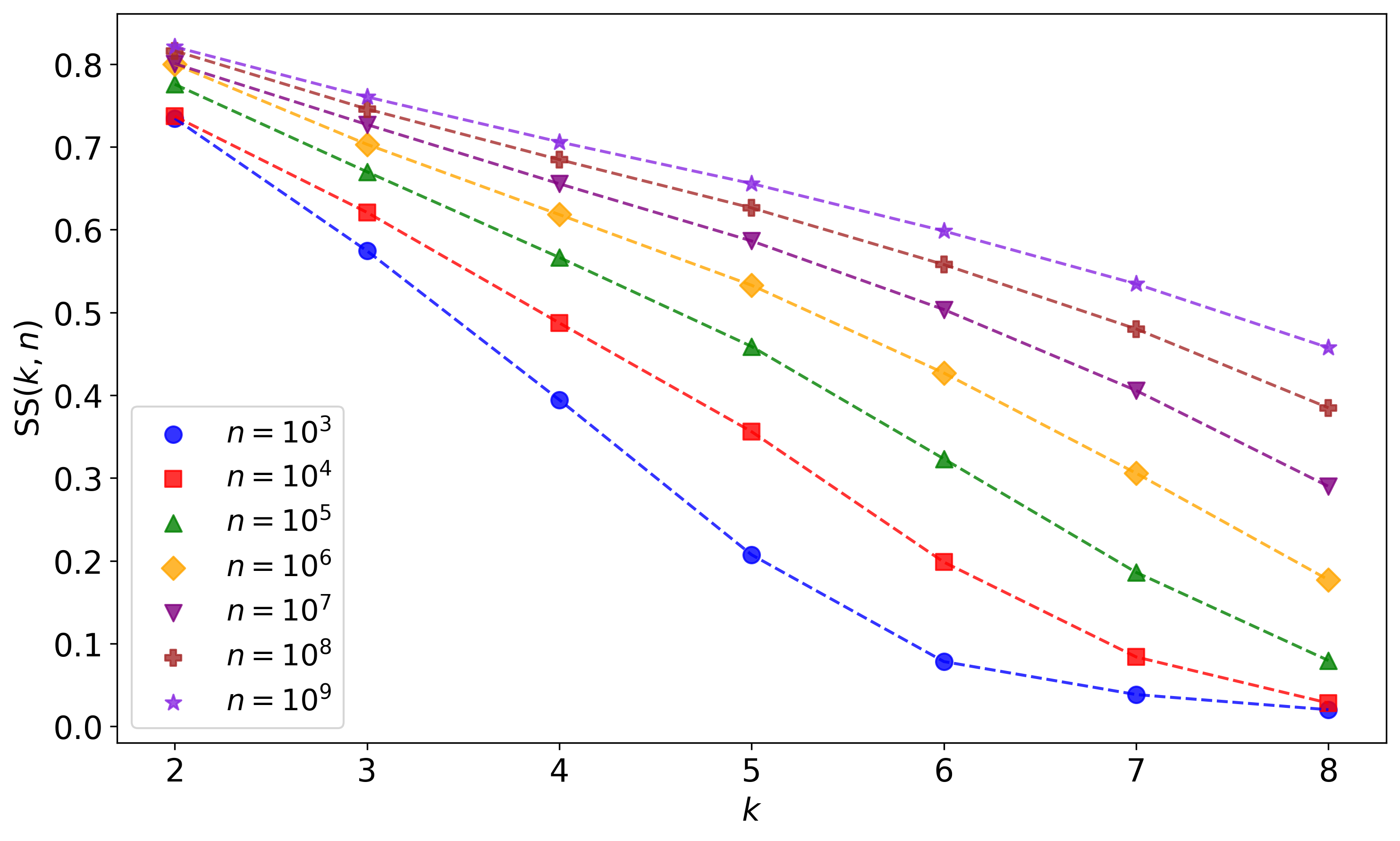}
\caption{Statistical Symmetry ${\rm SS}(k,n)$, defined in  \eqref{eq:defSym}, as a function of the tuple size $k$ for several values of the length of ${\mathbb{N}\mathcal{T}}_{1}^{n}$.}
\label{mirror-sym}
\end{figure}

 
We have computed \eqref{eq:defSym} for the length of the $\mathbb{N}$atural Text ${\mathbb{N}\mathcal{T}}_{1}^{n}$ ranging from $n=10^3$ to $n=10^9$ and for $k$-tuples with $k=2,\ldots,8$, see Figure \ref{mirror-sym}. 
%The results show that, for fixed $k \ge 3$, ${\rm Sym}(k,n)$ is increasing in $n$ and decreasing in $k$ for fixed $n$. In any case we have that the asymmetry is well below 1, since $ {\rm Sym}(k,n) \le  {\rm Sym}(k,10^8)<1$, suggesting that the text is clearly oriented. 
The results show that, for fixed $n$,  ${\rm SS}(k,n)$ is decreasing in $k$,  which means that the longer the subsequence, the smaller the chance of finding its mirror-reversed version within the given window of the \Natv. 
On the other hand, within the range of $n$ values we have considered, the statistical symmetry ${\rm SS}(k,n)$ increases with $n$ for fixed $k$. This fact shows that the Statistical Symmetry is well below $1$, since $ {\rm SS}(k,n) \le  {\rm SS}(k,10^9)<0.821$, demonstrating that the text ${\mathbb{N}\mathcal{T}}_{1}^{10^9}$ 
is oriented, at least with regard to subsequences that are not too long ($k\le 8$). For example, more than $29\%$ of the sequences of length $4$ and more than $54\%$ of those of length $8$ are non-invertible in ${\mathbb{N}\mathcal{T}}_{1}^{10^9}$. This suggests the existence of correlations between Dyck words, which will be examined in Section \ref{sez:MSD}.

%\subsection{Pattern analysis in tree encoding}

\subsection{Complexity via Entropy and Compression}\label{sez:CC}
The fact that the dictionary density is decreasing approximately as $\delta_n \sim n^{-0.7656}$ for $n\le \overline{N}$ (see Section \ref{sez:DD}) implies that the number of distinct Dyck words, i.e. the dictionary size $|{\mathbb{N}\mathcal{D}}_{1}^{n}|$, is significantly smaller than the total number of Dyck words in the $\mathbb{N}$atural Text ${\mathbb{N}\mathcal{T}}_{1}^{n}$  of length $n \le \overline{N}$. As a consequence, some Dyck words are bound to appear more than once in the texts, and we are interested in computing the multiplicities of these occurrences. For this purpose, we introduce the {\em empirical frequency} of the Dyck word $\rho \in {\mathbb{N}\mathcal{D}}_{1}^{n}$ which is defined as
\begin{equation}\label{eq:emp-freq}
    p(\rho; {\mathbb{N}\mathcal{T}}_{1}^{n}) = \frac{1}{n} \sum_{i=1}^n  \mathbf{1}\! \big(\rho =x_i \big),
\end{equation}
%p(n,\rho)=\frac{1}{n} \sum_{i=1}^n  \mathbf{1}\! \big(\rho =x_i \big)$ 
where $x_i$ is the $i$-th Dyck word of the \Nat ${\mathbb{N}\mathcal{T}}_{1}^{n}$. 
Thus, for instance, with reference to the text ${\mathbb{N}\mathcal{T}}_{1}^{10}$ given in equation \eqref{eq:nattext-1-10}, if $\rho=\texttt{10}$, we have that $p(\texttt{10}; {\mathbb{N}\mathcal{T}}_{1}^{10})=\frac{4}{10}$, while for $\rho=\texttt{1010}$ we have $p(\texttt{1010}; {\mathbb{N}\mathcal{T}}_{1}^{10})=\frac{2}{10}$.\par
Given the frequencies of all Dyck words in the dictionary ${\mathbb{N}\mathcal{D}}_{1}^{n}$, we treat them as a probability distribution ${\mathcal{P}_n}=(p(\rho; {\mathbb{N}\mathcal{T}}_{1}^{n}); \rho \in {\mathbb{N}\mathcal{D}}_{1}^{n})$ from which we can compute the Shannon entropy of the \Nat ${\mathbb{N}\mathcal{T}}_{1}^{n}$:
\begin{equation}\label{eq:def-entropy}
h({\mathbb{N}\mathcal{T}}_{1}^{n})=- \sum_{\rho \in {\mathbb{N}\mathcal{D}}_{1}^{n}}  p(\rho; {\mathbb{N}\mathcal{T}}_{1}^{n}) \log_2 \,  p(\rho; {\mathbb{N}\mathcal{T}}_{1}^{n}),
\end{equation}
that quantifies the amount of uncertainty (or information) contained in ${\mathbb{N}\mathcal{T}}_{1}^{n}$. %In \eqref{eq:def-entropy} $\log$ denotes $\log_2$.
Since the entropy of a $m$-component probability vector is not larger than $\log_2 m$, using the estimate $|{\mathbb{N}\mathcal{D}}_{1}^{n}|\sim K n^\beta$ given in Section \ref{sez:DD}, we can compute an {\em entropic bound} for our \Nat, as follows:
\begin{equation} \label{eq:bondnetropy}h({\mathbb{N}\mathcal{T}}_{1}^{n})\le   3.207 + 0.234\log_2 n
\end{equation}
for  $n\le 6.5\cdot 10^9$.\par
\begin{figure}[H]
    \centering
\includegraphics[width=13cm]{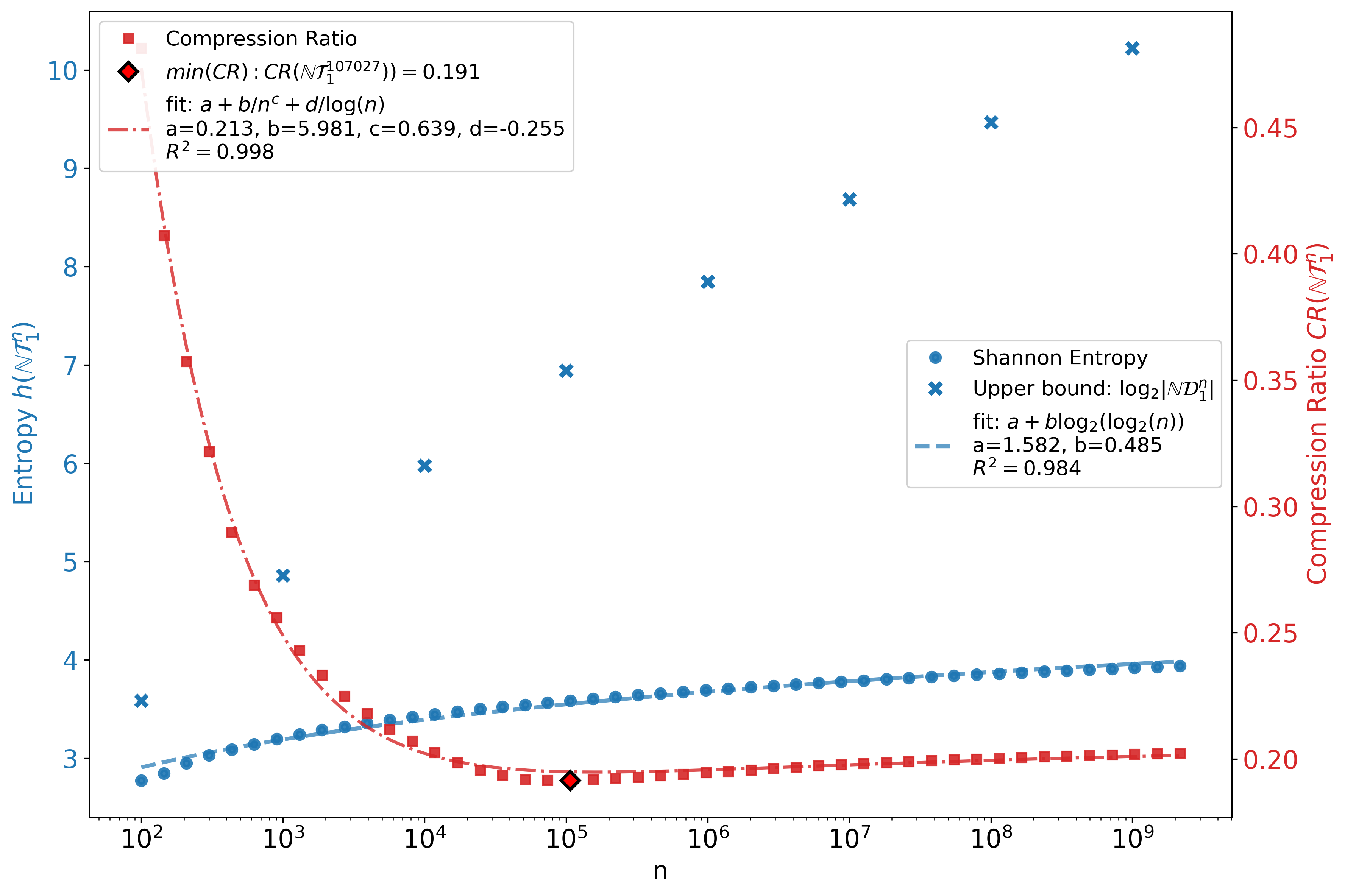}
\caption{Empirical entropy (blue dots) and compression ratio (red squares), as functions of sequence length $n$, together with model fits. Blue crosses represent the theoretical upper bound of entropy, given by $\log_2 |{\mathbb{N}\mathcal{D}}_{1}^{n}|$. The diamond marks the point where the compression rate reaches its minimum.}
    \label{fig:entropy-comp-min}
\end{figure}

Figure \ref{fig:entropy-comp-min},  where $h({\mathbb{N}\mathcal{T}}_{1}^{n})$ is represented as a function of $n$ (blue bullets), 
shows that the entropy is closer to its maximum possible value for small $n$, but significantly lower for large $n$.
In fact, the growth of entropy in the \Nat ${\mathbb{N}\mathcal{T}}_{1}^{n}$ reveals a distinct scaling behaviour relative to its theoretical maximum. Although the empirical entropy increases from approximately $2.77$ bits (at $n=10^2$) to $3.94$ bits (at $n\approx 2.16 \cdot10^9$), it remains substantially below the corresponding upper bounds (see equation \eqref{eq:bondnetropy}), which range from $3.58$ to $10.22$ bits on the same scale. This persistent gap demonstrates that ${\mathbb{N}\mathcal{T}}_{1}^{n}$ possesses strong redundancy and a highly non-uniform distribution of Dyck words, indicating significant inherent structure rather than randomness.\par

It is appropriate to make some considerations regarding the computation of entropy and its validity as $n$ varies. Although the evaluation of entropy of $\mathcal{P}_n$ is virtually exact for fixed $n$ since the empirical frequencies can be computed precisely (up to rounding errors), we clearly cannot extrapolate our results beyond the maximum length $\overline{N}$ of the window of the \Nat available to us. 
It is well known  \cite{SchuGRas} that the estimation of the entropy can be problematic even in `standard' contexts — namely, for sequences with a finite number of distinct symbols occurring at well-defined and stationary frequencies. Indeed, the presence of complex and long-range correlations makes the evaluation of entropy nontrivial, since \eqref{eq:emp-freq} and \eqref{eq:def-entropy} generally underestimate this quantity due to the large fluctuations that may occur  in \eqref{eq:emp-freq} as $n$ varies. In our case, the estimation is even more difficult, as we are dealing with a non-standard situation as evidenced, for instance, by the fact that Dyck word frequencies vanish as 
$n$ becomes larger and larger.

To quantify the observed redundancy, we compute the compression ratio of ${\mathbb{N}\mathcal{T}}_{1}^n$ using the \texttt{gzip} algorithm \cite{Ziv_Lempel_1977}: %under the UTF-8 encoding scheme

%\begin{equation}
%R({\mathbb{N}\mathcal{T}}_{1}^{n}) \;=\; %\frac{|C({\mathbb{N}\mathcal{T}}_{1}^{n})|}%{S({\mathbb{N}\mathcal{T}}_{1}^{n})}
%\end{equation}
\begin{equation}
CR({\mathbb{N}\mathcal{T}}_{1}^{n}) \;=\; \frac{C({\mathbb{N}\mathcal{T}}_{1}^{n})}{S({\mathbb{N}\mathcal{T}}_{1}^{n})}
\end{equation}
where $C({\mathbb{N}\mathcal{T}}_{1}^{n})$ denotes the file size of the compressed representation of  ${\mathbb{N}\mathcal{T}}_{1}^n$ and $S({\mathbb{N}\mathcal{T}}_{1}^{n})$ its original file size. The values of $CR({\mathbb{N}\mathcal{T}}_{1}^{n}) $ closer to $0$ indicate higher compressibility and redundancy, whereas those closer to $1$ correspond to incompressible or random-like data.

The compression ratio (red squares in Figure \ref{fig:entropy-comp-min}) decreases rapidly for small $n$, possibly reflecting the progressive discovery of some underlying rules and regularities associated with the natural number factorisation.  %its recursive regularities and subtree motifs associated with integer factorisation. 
This phase corresponds to the identification of local redundancy and short-range correlations. A minimum is reached around $n = 10^5$, where the representation achieves maximal compactness, indicating that syntactic regularities are most efficiently captured. Beyond this point, the mild increase in $CR({\mathbb{N}\mathcal{T}}_{1}^{n})$ suggests the emergence of higher-level variability in which new, less frequent tree configurations and long-range dependencies reduce redundancy. A fit of the function $a+b/n^c + d/\log n$ to the data (which should not be considered asymptotic) highlights the slow growth of $CR({\mathbb{N}\mathcal{T}}_{1}^{n}) $, see Figure \ref{fig:entropy-comp-min}.

The entropy trend corroborates this interpretation. Although $h({\mathbb{N}\mathcal{T}}_{1}^{n})$ increases, it remains significantly below its theoretical upper bound (blue crosses), confirming that the sequence is far from random. The growth of entropy is extremely slow, as can be appreciated from the fit with $a+b\log_2(\log_2(n))$ shown in Figure  \ref{fig:entropy-comp-min} (once again, we emphasize that the fit is not intended to represent an asymptotic estimate).
This sublinear growth implies that additional Dyck words contribute diminishing information per symbol, consistent with a constrained but generative structure.

Taken together, the entropy and compression analyses delineate a transition from order to organized complexity. The \Nat, while growing, maintains an internal generative hierarchy that continually introduces novel configurations without losing coherence. The coexistence of low entropy and partial loss of compressibility characterise a regime of structured complexity, in which predictability and variability coexist in what appears statistically coherent and self-organizing form.

\subsection{The Rank-Frequency Distribution of Dyck words: the Zipf Function}
\begin{figure}[htbp]
    \centering
    \includegraphics[width=0.9\textwidth]{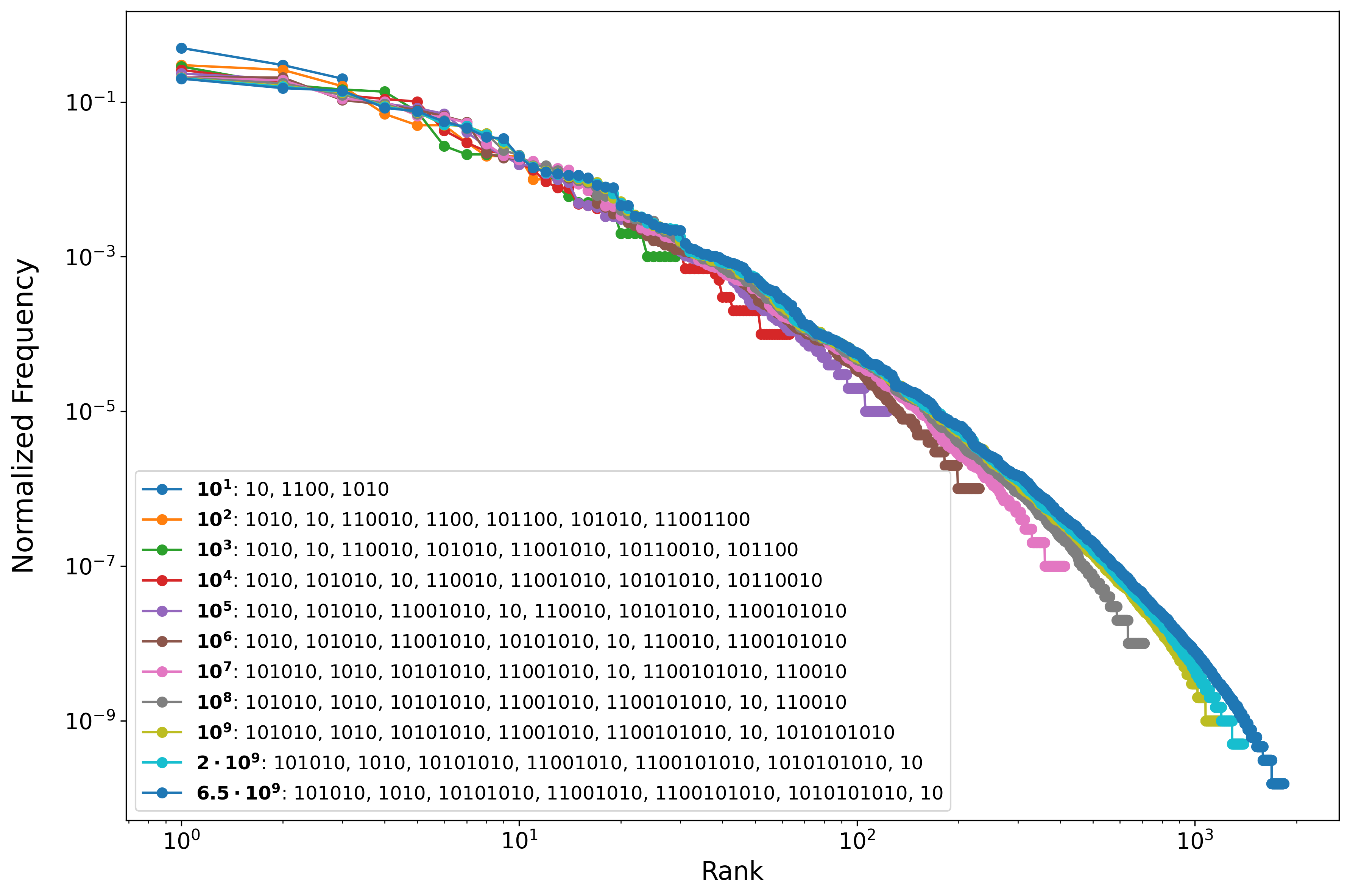}
 \caption{Log-log plot of the rank-frequency distributions $\mathcal{F}_n$ (i.e. Zipf functions $f_n(r)$) versus the rank $r$,  for $n$ varying from $10$ up to $6.5\cdot 10^9$. In the legend, for each $n$, the most frequent Dyck words are reported.}
    \label{zipf-norrmalized}
\end{figure}

%To further characterise the statistical organization of the $\mathbb{N}$atural text, we examine the distribution of its empirical frequencies. 
%In this section we discuss the frequency-rank distribution of the $\mathbb{N}$atural text using Zipfian and non-Zipfian arguments.
In this section, to further characterise the statistical properties of the \Natv, we examine the {\em rank-frequency distribution} $\mathcal{F}_n=(f_n(r),\, r=1,\ldots, d_n)$ of the empirical vector ${\mathcal{P}_n}$ whose components are given in \eqref{eq:emp-freq} and whose dimension is $d_n$. The components $f_n(r)$ of  $\mathcal{F}_n$ are obtained by rearranging those of ${\mathcal{P}_n}$ in decreasing order. More explicitly: $f_n(1)\ge f_n(2)\ge \cdots \ge f_n(r)\ge \cdots$ are the frequencies appearing in ${\mathcal{P}_n}$ and $r=1,2,\ldots$ are the corresponding {\em ranks}. Figure \ref{zipf-norrmalized} displays the function $f_n(r)$, called {\em Zipf function}, for values of $n$ ranging from $10$ to $\overline{N}$. The figure highlights a remarkable fact 
%that they are, to a good approximation, independent of $n$.
that {\em the Zipf function $f_n(r)$ is, to a good approximation,  independent of $n$}. 
Thus, for example, in every text ${\mathbb{N}\mathcal{T}}_{1}^{n}$ of length $n\le \overline{N}$ the most frequent Dyck word, i.e. the one of rank 1, has an approximate frequency of $20\%$, while the frequency of that of rank 2 is approximately $15\%$.\par
We stress that this analysis shows that {\em while the frequencies \eqref{eq:emp-freq} of individual Dyck words vary with $n$ %and are not well-defined in the limit of large $n$
 (see the discussion in Section \ref{sez:CC} ), the rank frequencies tend to remain approximately constant with respect to $n$. This suggests that the rank frequencies may be well-defined even in the limit as $n$ tends to infinity.
Moreover, it may also be worth noting that the invariance of the rank-frequency distributions with respect to $n$ does not imply the invariance of the ranks of individual Dyck words.}\par
That is, denoting $\rho_n(r)$  the Dyck word of rank $r$ in ${\mathbb{N}\mathcal{T}}_{1}^{n}$, one can check that this Dyck word can change as $n$ varies. As an example, Table 
\ref{tab:rank-freq} shows how the five most frequent Dyck words in ${\mathbb{N}\mathcal{T}}_{1}^{n}$, $\rho_n(1), \rho_n(2), ..., \rho_n(5)$, change as $n$ changes. 
Despite the fact that the table further suggests that 
$\rho_n(1)$ and $\rho_n(2)$ are constant for $10^7\le n \le 6.5\cdot10^9$, as previously noted in other sections of this article, no definitive conclusions can be drawn regarding the asymptotic behaviour of the maps $\rho_n(r)$ in the limit as $n$ tends to infinity.\par

%Table \ref{tab:rank-freq}  further indicates that $\rho_n(1)$ and $\rho_n(2)$ are constant for $10^7\le n \le 2\cdot10^9$, however,  as already noted several times in this article, based on our data we cannot infer the behaviour of the maps $\rho_n(r)$ in the limit as $n$ tends to infinity.\par

We now examine in greater detail the nature of the Zipf function. A very common behaviour for the rank-frequency distribution encoded in the Zipf function $f(r)$, is described by the so called {\em Zipf's law}, in which the frequency $f(r)$ is inversely proportional to the rank $r$. Figure \ref{zipf-norrmalized} clearly shows that this law does not apply to our dataset. Instead, the decay exhibits a persistent curvature in $\log$–$\log$ plot, forming a scale independent parabolic envelope that extends over more than three orders of magnitude in rank. This behaviour is well approximated by another distribution, the {\em Parabolic Fractal Distribution} \cite{Laherrere1996, Laherrere1998a}, according to which the logarithm of the frequency is a quadratic function of the logarithm of the rank. 

\begin{figure}[H]
    \centering
\includegraphics[width=0.8\textwidth]{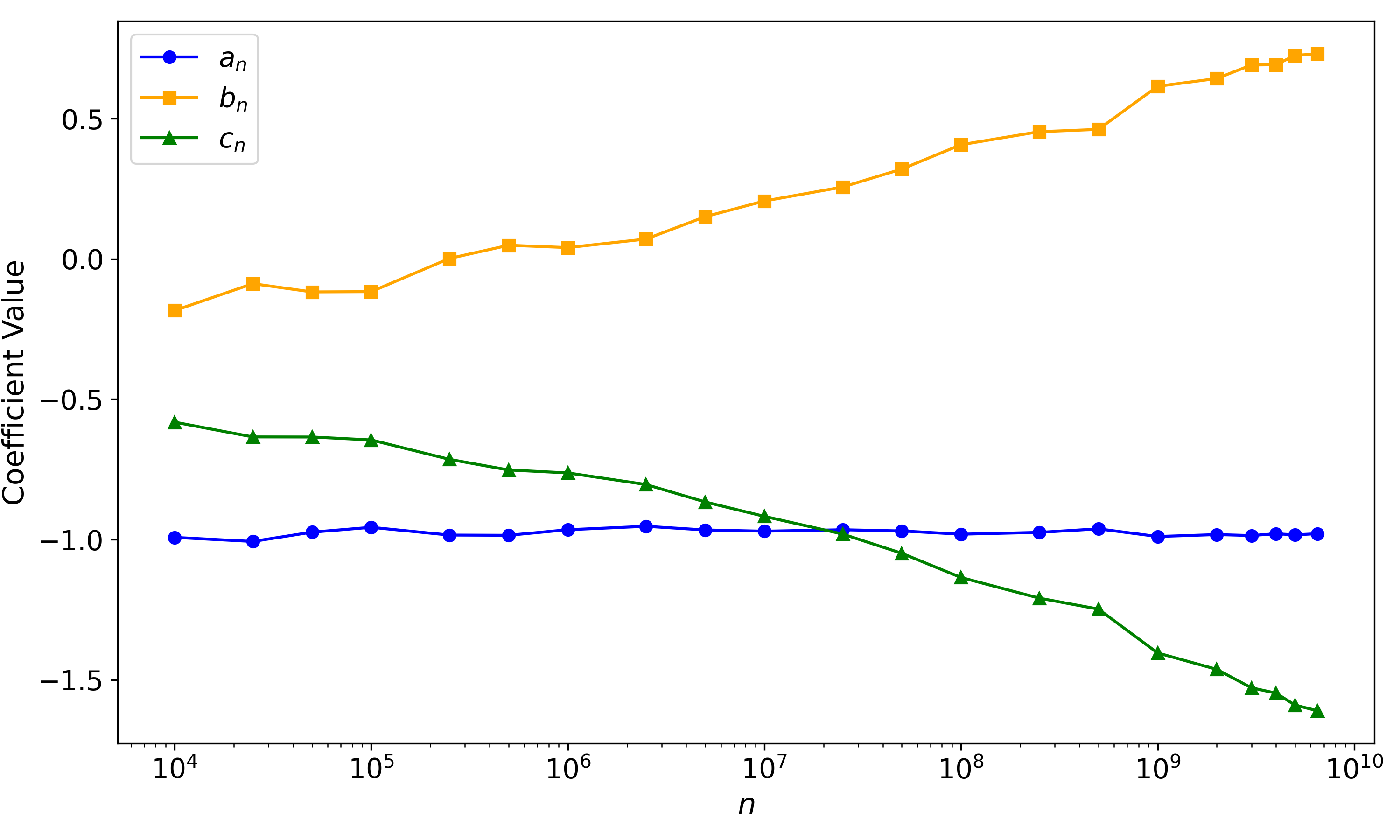}
\caption{Coefficients of the quadratic fits \eqref{eq:pfd} for the rank-frequency distributions $\mathcal{F}_n$ (Zipf functions) versus $n$ varying from $10^4$ to 6.5$\cdot 10^9$.  The coefficients $a_n$, $b_n$ and $c_n$ are representd by blue dots, orange squares and green triangles, respectively.}
    \label{fig:quadratic fit}
\end{figure}

We have fitted the rank-frequency distributions $\mathcal{F}_n$  to
\begin{equation}\label{eq:pfd}
\log f_n(r) = a_n\,(\log r)^{2} + b_n\,\log r + c_n
\end{equation}
for $n$ ranging from $10^{4}$ to $6.5\cdot 10^{9}$. Figure \ref{fig:quadratic fit}, in which the coefficients of the fits are reported versus $n$, shows that the quadratic coefficient $a_n$ remains remarkably stable around $-1$ across all scales, corroborating the observation that the distributions $\mathcal{F}_n$ are almost independent of $n$, at least with respect to the dominant term $(\log r)^2$ in \eqref{eq:pfd}.

%For instance, denoting $\rho_n(r)$  the Dyck word of rank $r$ in ${\mathbb{N}\mathcal{F}}_{1}^{n}$,  we have: $\rho_{10}(1)=10$, $\rho_n(1)=1010$ for $n=10^2, 10^3, 10^4, 10^5, 10^6$ and $\rho_n(1)=101010$ for $n=10^7, 10^8, 10^9, 2\cdot 10^9, 6.5\cdot 10^9$.
\begin{table}[]
    \centering
    \begin{tabular}{c|c|c|c|c|c}
        $n$ & $\rho_n(1)$ &  $\rho_n(2)$ & $\rho_n(3)$ & $\rho_n(4)$ &  $\rho_n(5)$ \\
        \hline
        10 & \texttt{10} &  \texttt{1100}    & \texttt{1010}  & -- & --  \\
        $10^2$& \texttt{1010} & \texttt{10} & \texttt{110010} & \texttt{1100} & \texttt{101100} \\
        $10^3$& \texttt{1010} & \texttt{10} & \texttt{110010} & \texttt{101010} &  \texttt{11001010} \\
        $10^4$ & \texttt{1010} &  \texttt{101010} & \texttt{10} &  \texttt{110010} & \texttt{11001010} \\
        $10^5$ & \texttt{1010} &  \texttt{101010} & \texttt{11001010} & \texttt{10} &  \texttt{110010} \\
        $10^6$ & \texttt{1010} &  \texttt{101010} & \texttt{11001010} & \texttt{10101010} & \texttt{10}  \\
        $10^7$ & \texttt{101010} & \texttt{1010}  & \texttt{10101010} & \texttt{11001010} & \texttt{10} \\ 
        $10^8$ & \texttt{101010} &  \texttt{1010} & \texttt{10101010} & \texttt{11001010} & \texttt{1100101010}  \\ 
        $10^9$ & \texttt{101010} &  \texttt{1010} & \texttt{10101010} & \texttt{11001010} & \texttt{1100101010}   \\ 
        $2\cdot 10^9$& \texttt{101010} & \texttt{1010} & \texttt{10101010} & \texttt{11001010} & \texttt{1100101010}  \\
        $6.5 \cdot 10^9$& \texttt{101010} & \texttt{1010} & \texttt{10101010} & \texttt{11001010} & \texttt{1100101010} 
    \end{tabular}
    \caption{Dyck words of rank 1, $\rho_n(1)$, to rank 5, $\rho_n(5)$, in ${\mathbb{N}\mathcal{T}}_{1}^{n}$ for several values of $n$.}
    \label{tab:rank-freq}
\end{table}

\subsection{Correlation via Mean Squared Displacement}\label{sez:MSD}
In this section, we continue to study the distribution of occurrences of the Dyck words of the \Nat by using quantities such as the mean square displacement and the correlation functions, which are commonly employed  in  signals or stochastic processes analysis \cite{Michalet_2010} or in the study of natural languages; see, e.g. \cite{Manin_2008, AltCriDesp}. \par
Given a Dyck word $\rho \in \mathbb{N}\mathcal{D}$, we construct the sequence ${\bf z}=(z_i)_{i=1}^n$ (the ``signal")  whose components are
\begin{equation}\label{eq:zeoone}
z_i = \begin{cases}
1 & \text{if } x_i = \rho \\
0 & \text{othewise}
\end{cases}
\end{equation}
where $x_i$ is the $i$-th Dyck word of the text  ${\mathbb{N}\mathcal{T}}_{1}^{n}$, see \eqref{eq:nattextx}.
In order to construct a walk based on ${\bf z}$, we introduce the centered sequence ${\bf y}$ with components $y_i=z_i-\mu$,
%So, for instance, the first 10 components of the sequence $\bf z$ for $a=10$ are
%$
%{\bf z}=(0,1,1,0,1,0,1,0,0,0,\ldots).
%$
where $\mu={\overline{N}}^{-1}  \sum_{i=1}^{\overline{N}} z_i$, and define the position of a detrended walker ${\bf s}$ at ``time'' $n$ as $s_n= \sum_{i=1}^n y_i$. For the ``temporal'' dispersion of this walk from step $n$ onward, we consider the displacement over a lag $t$:
$$
\Delta s_{n,t} = s_{n+t}-s_n= \sum_{i=n+1}^{n+t} y_i\,.
$$
Introducing the average $\langle \cdot \rangle$  over the \Natv, 
we compute the {\em mean square displacement} at time $t$ of the walk as:
$$
    {\rm MSD}_\rho (t)= \langle (\Delta s_{n,t})^2 \rangle = \frac{1}{\overline{N}-t} \sum_{n=1}^{\overline{N}-t} \left ( \sum_{i=n+1}^{n+t} y_i\right)^2\,.  
$$
%which reads as:
%\begin{equation}
%     {\rm MSQ}_\rho (t)= \frac{1}{N-t} \sum_{n=1}^{N-t} \left ( %\sum_{i=n+1}^{n+t} y_i\right)^2 
%\end{equation}
\invisible{In general, the time scaling of the mean square displacement can be written as
\begin{equation}
     {\rm MSD}_\rho (t) \sim t^\gamma\,,
\end{equation}
where the exponent $\gamma>0$ defines the possible transport regimes.} 
The mean square displacement typically exhibits a power-law behaviour,  
\begin{equation}
     {\rm MSD}_\rho (t) \sim t^\gamma\,,
\end{equation}
where the exponent $\gamma>0$ identifies  the possible behaviours of the walker.
In particular, according to the terminology of transport theory, we have that {\em normal diffusion} is characterised by $\gamma=1$, while in  anomalous cases we can recognize {\em subdiffusion} if $0<\gamma<1$, {\em superdiffusion} if  $1<\gamma<2$, and {\em ballistic} behaviour if $\gamma=2$. On the other hand, there are contexts in which transport is not characterised by a single exponent \cite{Riahi2019MSD, AwadMetzler2020}. This may occur in the presence of multiple temporal regimes as, for instance, in ageing continuous time random walks (CTRW), see e.g. \cite{Metzler2014},  in polymer dynamics \cite{Phillies2025Polymers} and in the behaviour of stock returns \cite{Ma2013EPL}. In such a scenario, the mean square displacement shows a {\em crossover} at some time $t_c$:
\begin{equation}
\label{eq:mad-crossover}
   {\rm MSD}_\rho (t) \sim 
   \left \{
   \begin{array}{lll}
   t^{\gamma_1} & \mbox{if} &t \ll t_c, \\
   t^{\gamma_2} & \mbox{if}  & t \gg t_c\,,
   \end{array}
   \right.
\end{equation}
with $\gamma_1 \ne \gamma_2$.
\par
\begin{figure}[H]
  \centering

  % Row 1
  \begin{subfigure}[b]{0.5\textwidth}
    \centering
    \includegraphics[width=\linewidth]{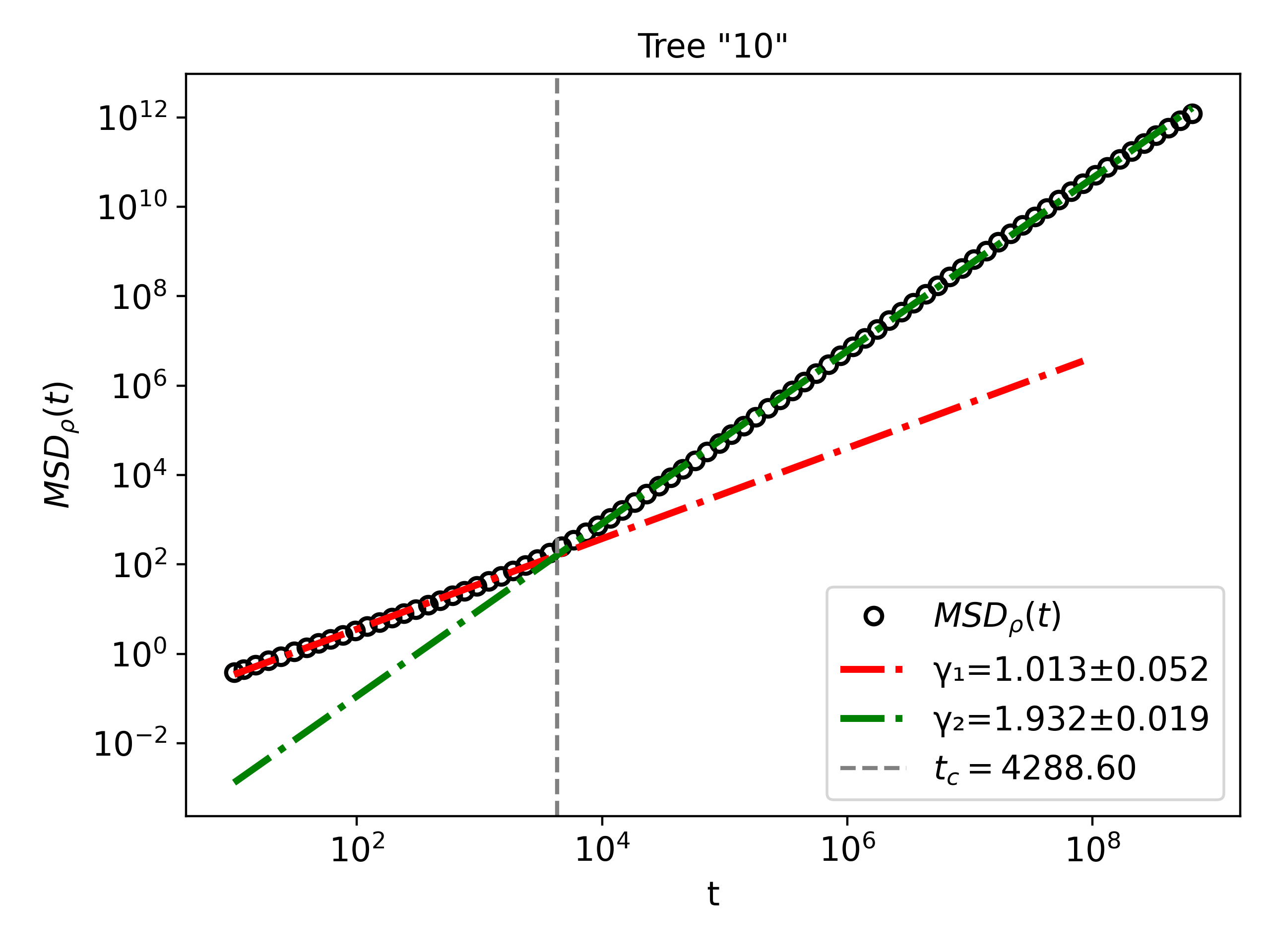}
    \label{MSD_gamma1}
  \end{subfigure}\hfill
  \begin{subfigure}[b]{0.5\textwidth}
    \centering
    \includegraphics[width=\linewidth]{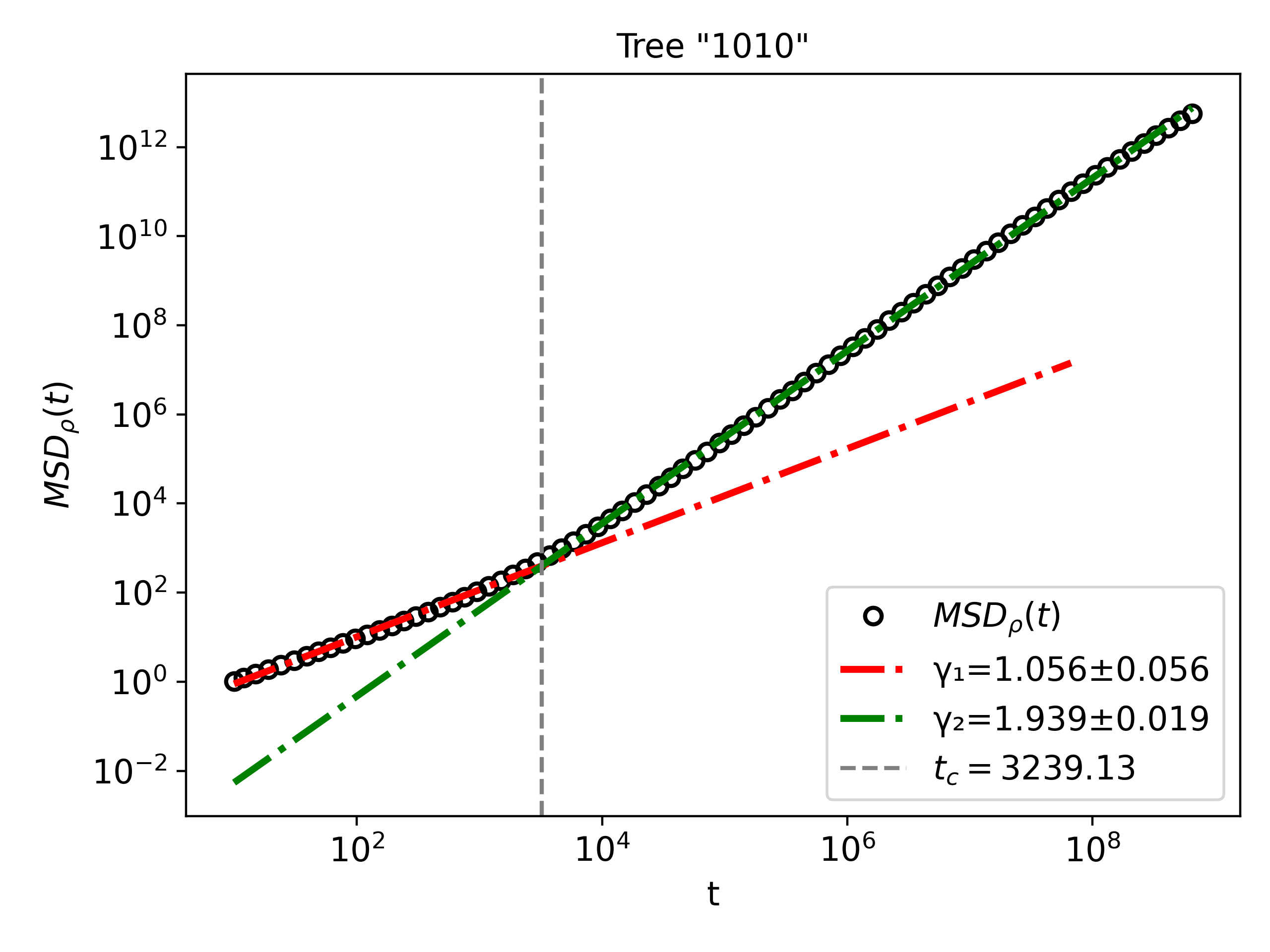}
    \label{MSD_gamm5}
  \end{subfigure}

  %\vspace{0.5em}

  % Row 2
  \begin{subfigure}[b]{0.5\textwidth}
    \centering
    \includegraphics[width=\linewidth]{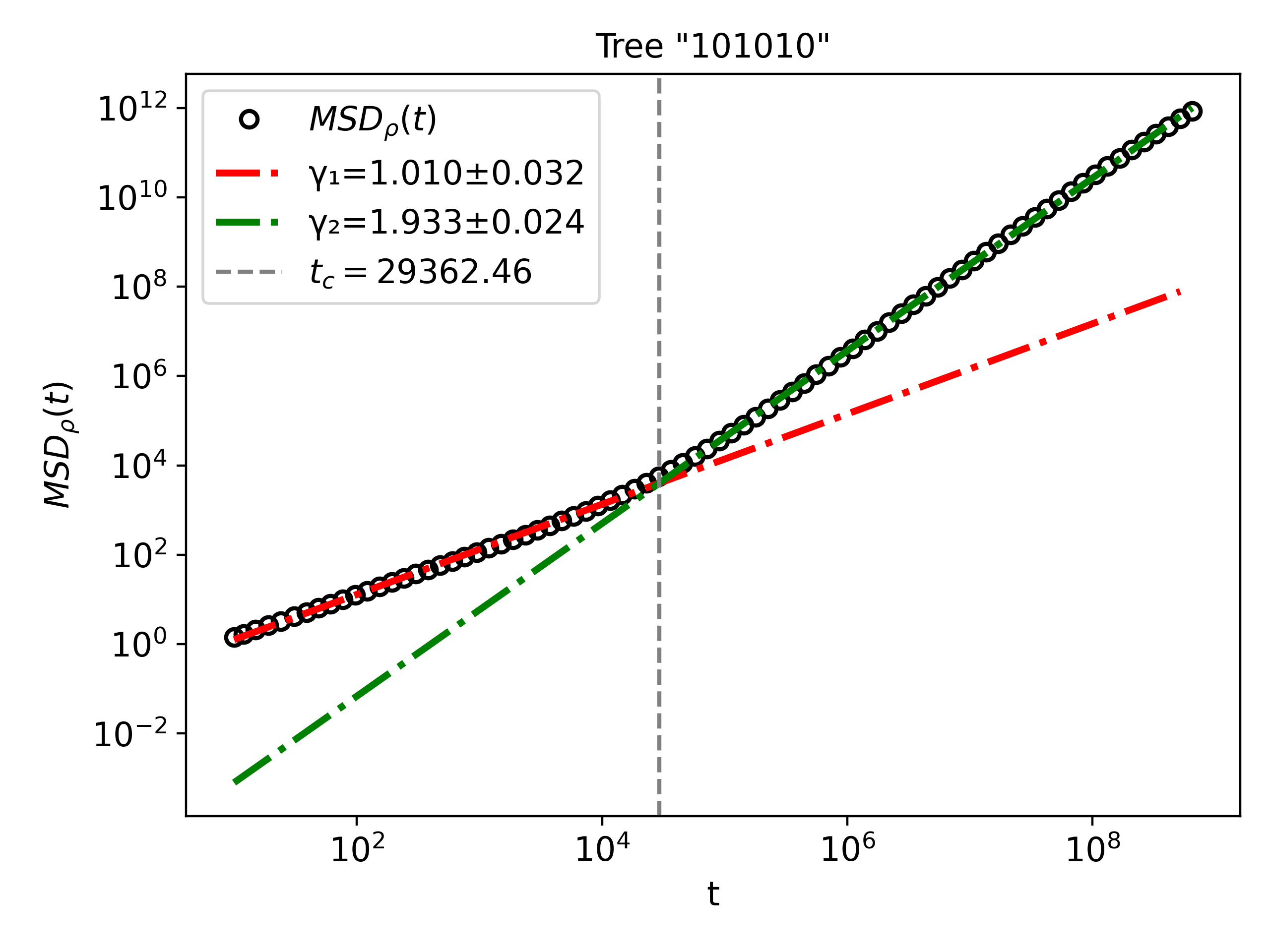}
    \label{MSD_gamma21}
  \end{subfigure}\hfill
  \begin{subfigure}[b]{0.5\textwidth}
    \centering
    \includegraphics[width=\linewidth]{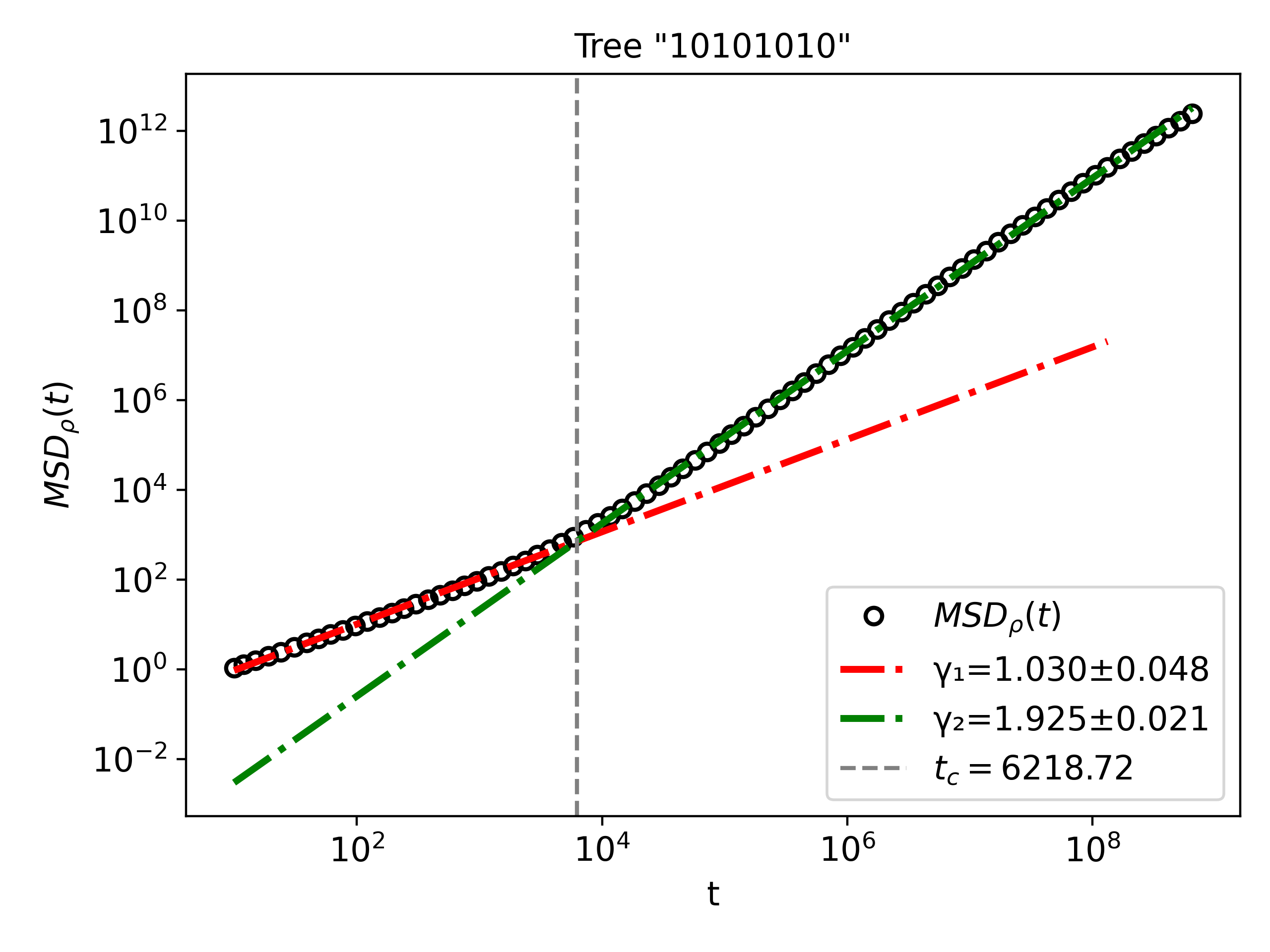}
    \label{MSD_gamm85}
  \end{subfigure}

  %\vspace{0.5em}

  % Row 3 (single)
  \begin{subfigure}[b]{0.5\textwidth}
    \centering
    \includegraphics[width=\linewidth]{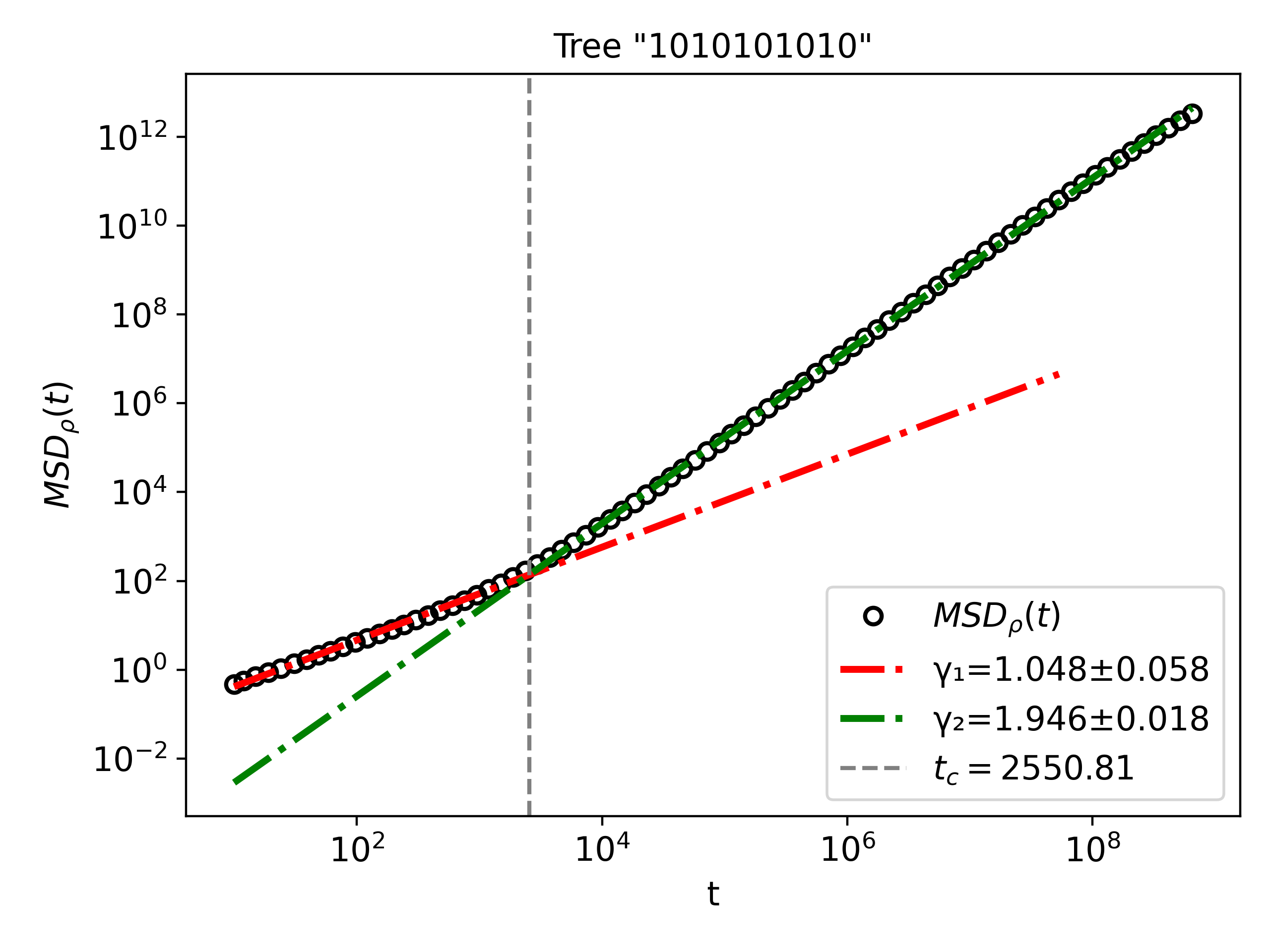}
    \label{MSD_gamma341}
  \end{subfigure}
  \caption{Log-log plots of mean square displacement $MSD_\rho(t)$ for Dyck word (Tree) $\rho$ associated with some square-free numbers versus the time lag $t$. The piecewise linear fits (dashed-dotted lines) illustrate  the phase changes from an approximately normal to a superdiffusive (quasi ballistic) behaviour. The vertical dashed line represents the crossover point $t_c$.}\label{msd_gamma_squarefree}
\end{figure}
\begin{figure}[H]
  \centering

  % Row 1
  \begin{subfigure}[b]{0.5\textwidth}
    \centering
    \includegraphics[width=\linewidth]{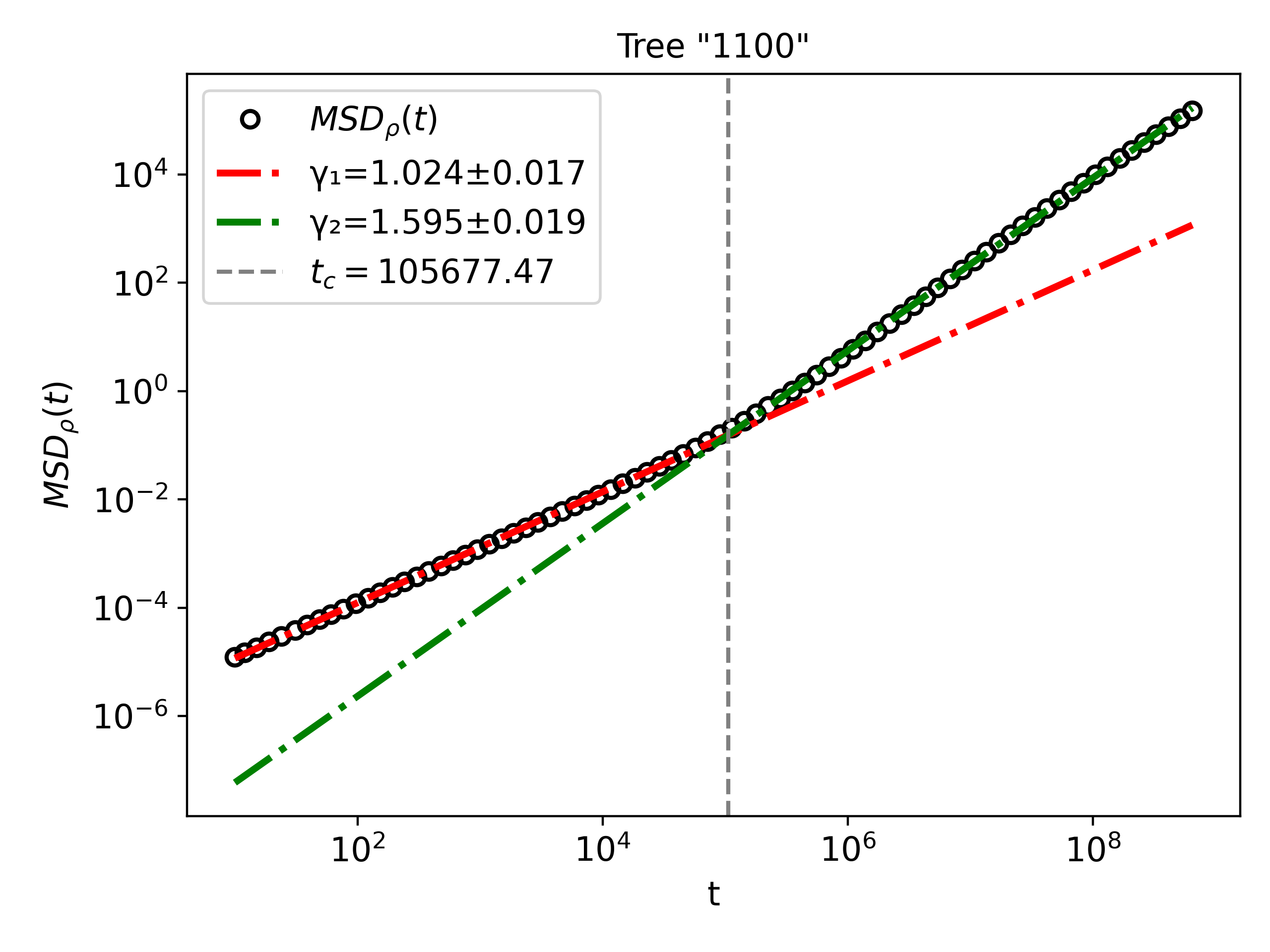}
\label{MSD_gamm6}
  \end{subfigure}\hfill
  \begin{subfigure}[b]{0.5\textwidth}
    \centering
    \includegraphics[width=\linewidth]{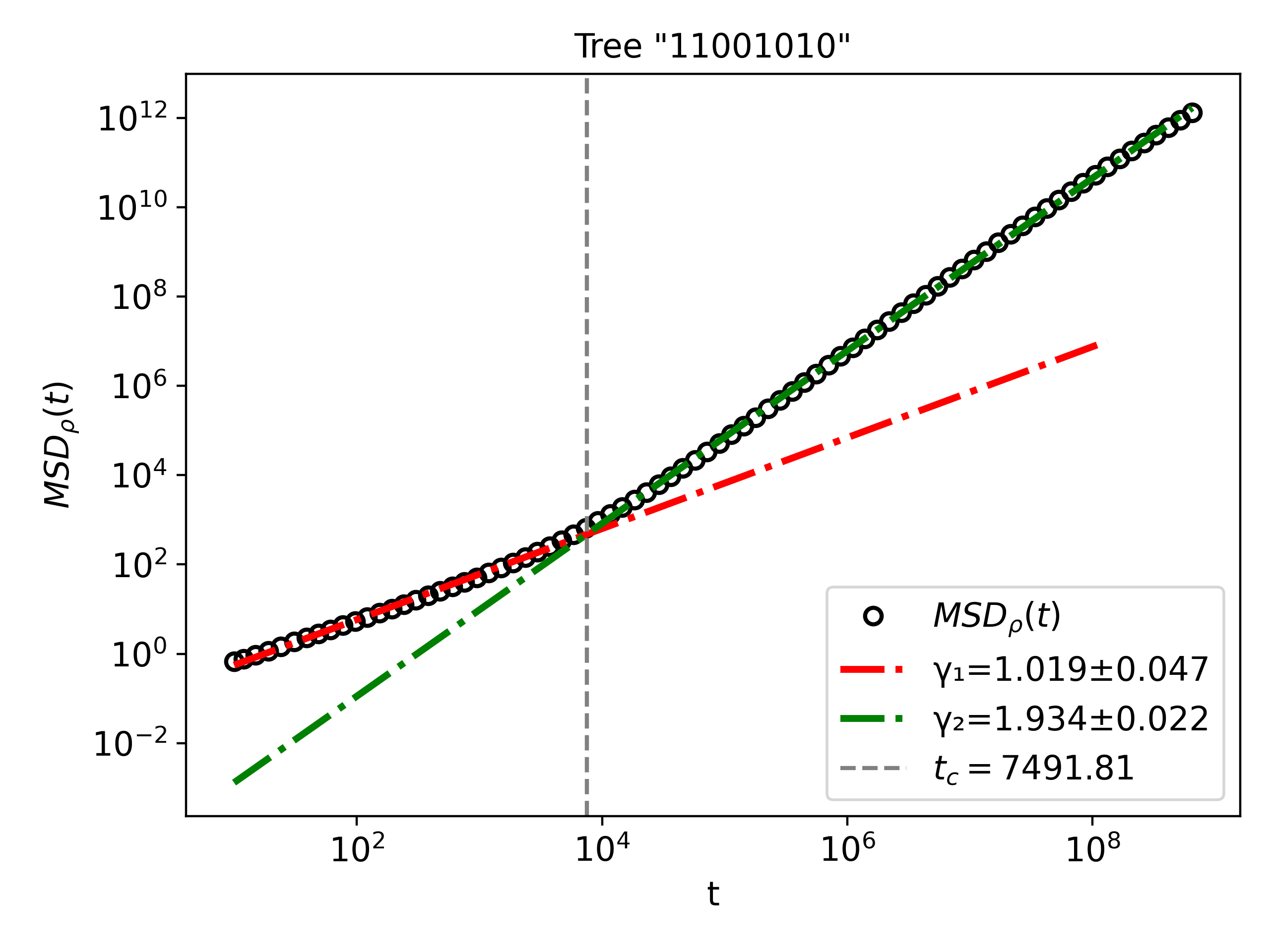}
\label{MSD_gamm101}
  \end{subfigure}

%  \vspace{0.5em}

  % Row 2
  \begin{subfigure}[b]{0.5\textwidth}
    \centering
    \includegraphics[width=\linewidth]{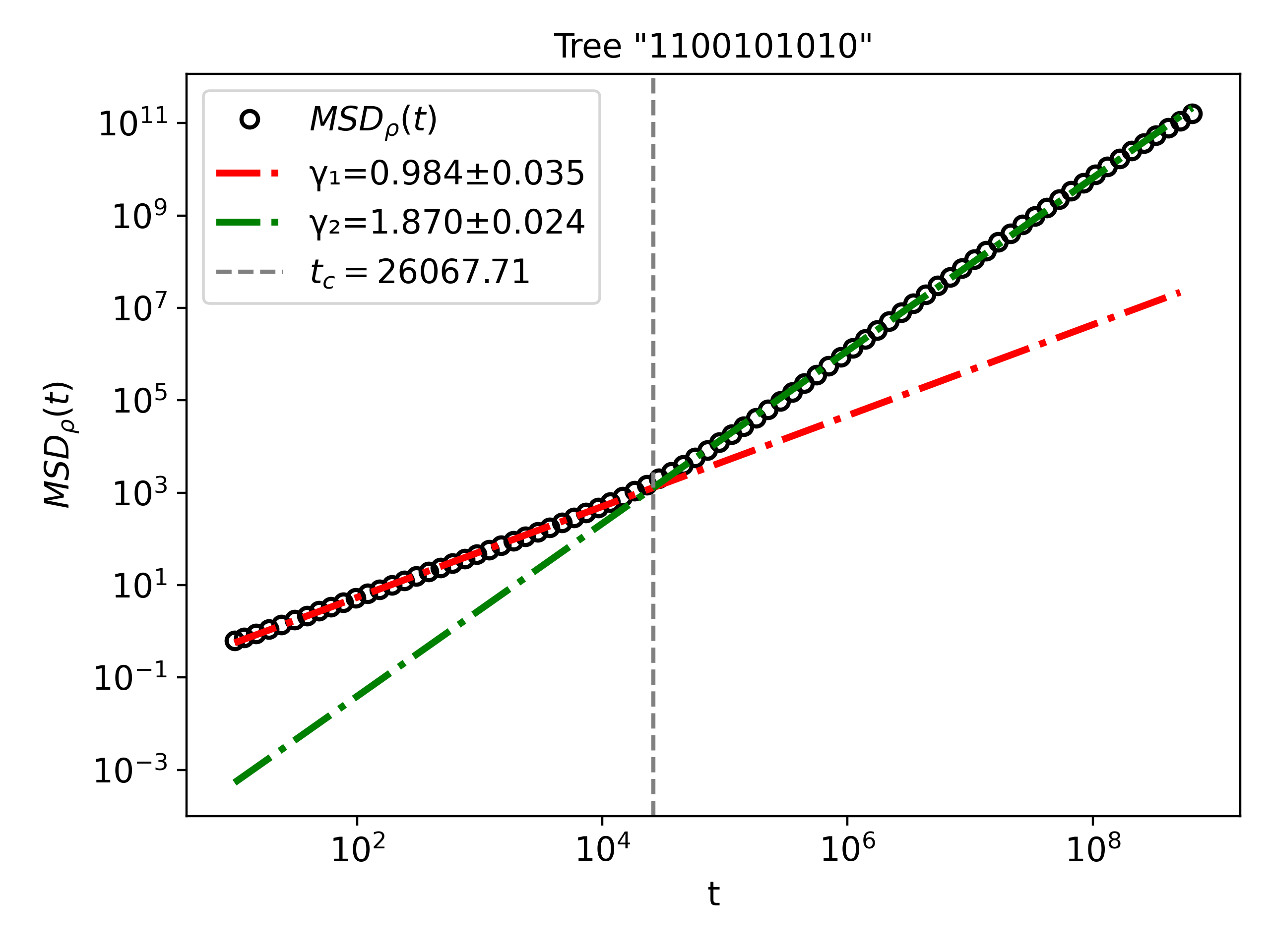}
\label{MSD_gamm405}
  \end{subfigure}\hfill
  \begin{subfigure}[b]{0.5\textwidth}
    \centering
    \includegraphics[width=\linewidth]{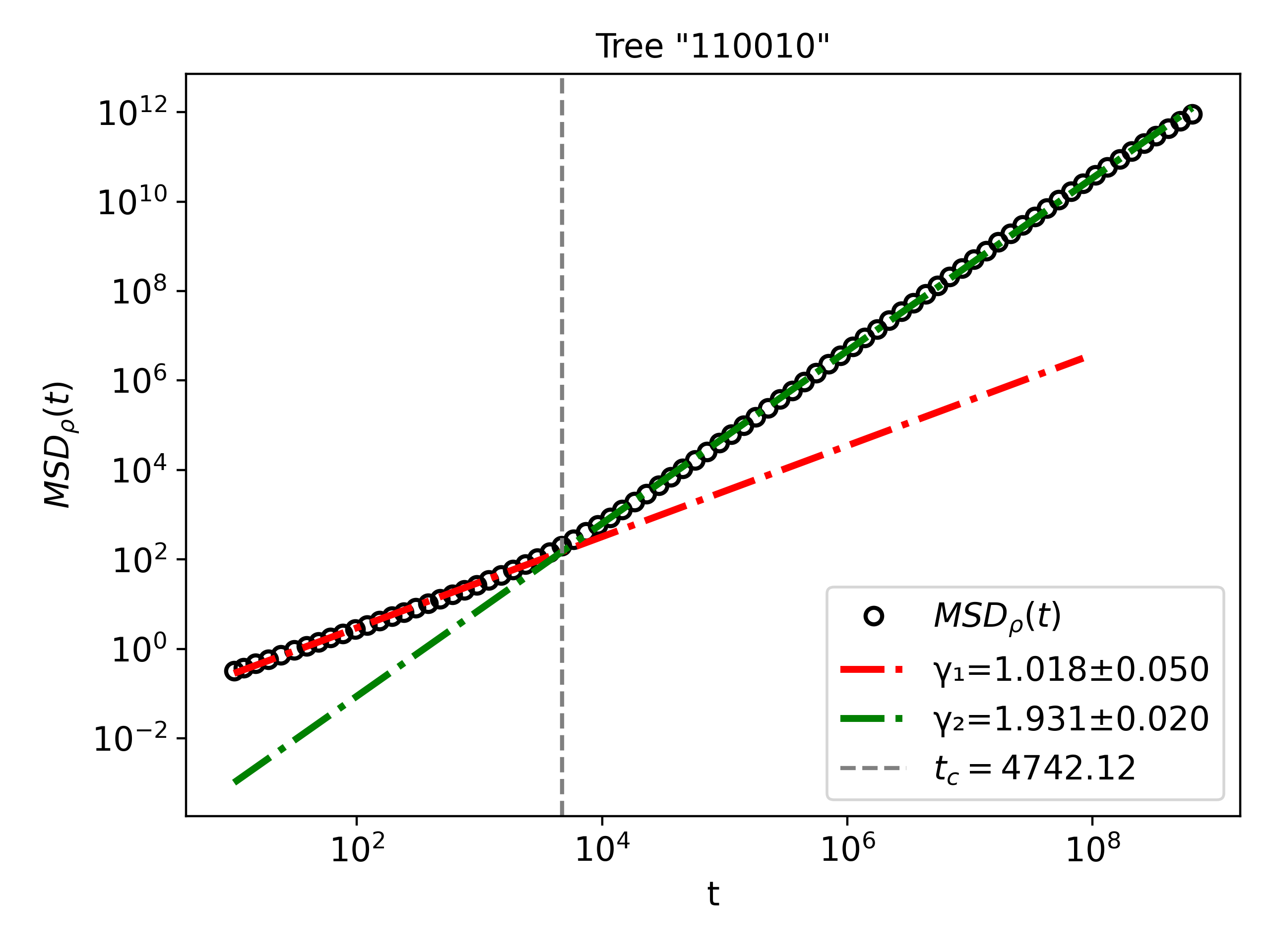}
\label{MSD_gamma25}
  \end{subfigure}

  %\vspace{0.5em}

  % Row 3 (single)
  \begin{subfigure}[b]{0.5\textwidth}
    \centering
    \includegraphics[width=\linewidth]{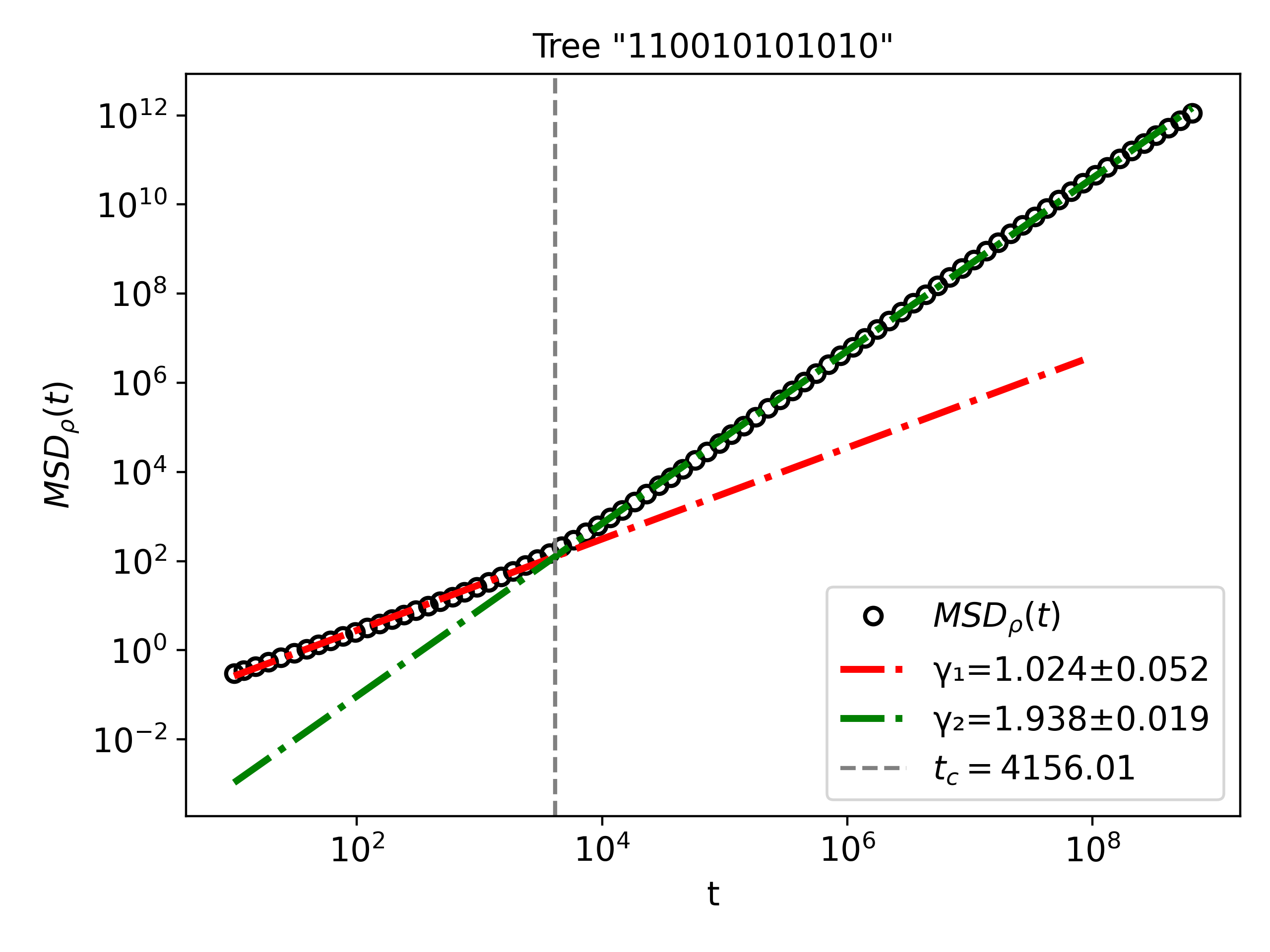}\label{MSD_gamm1621}
  \end{subfigure}
  \caption{Log-log plots of mean square displacement $MSD_\rho(t)$ for Dyck word (Tree) $\rho$ associated with some non-square-free numbers versus the time lag $t$. The piecewise linear fitting (dashed-dotted lines) illustrate  the phase changes from an approximately normal to a superdiffusive behaviour.  The vertical dashed line represents the crossover point $t_c$.}
  \label{msd_gammas_nonsquarefree}
\end{figure}
\begin{figure}[htbp]
  \centering
  \includegraphics[width=1\textwidth]{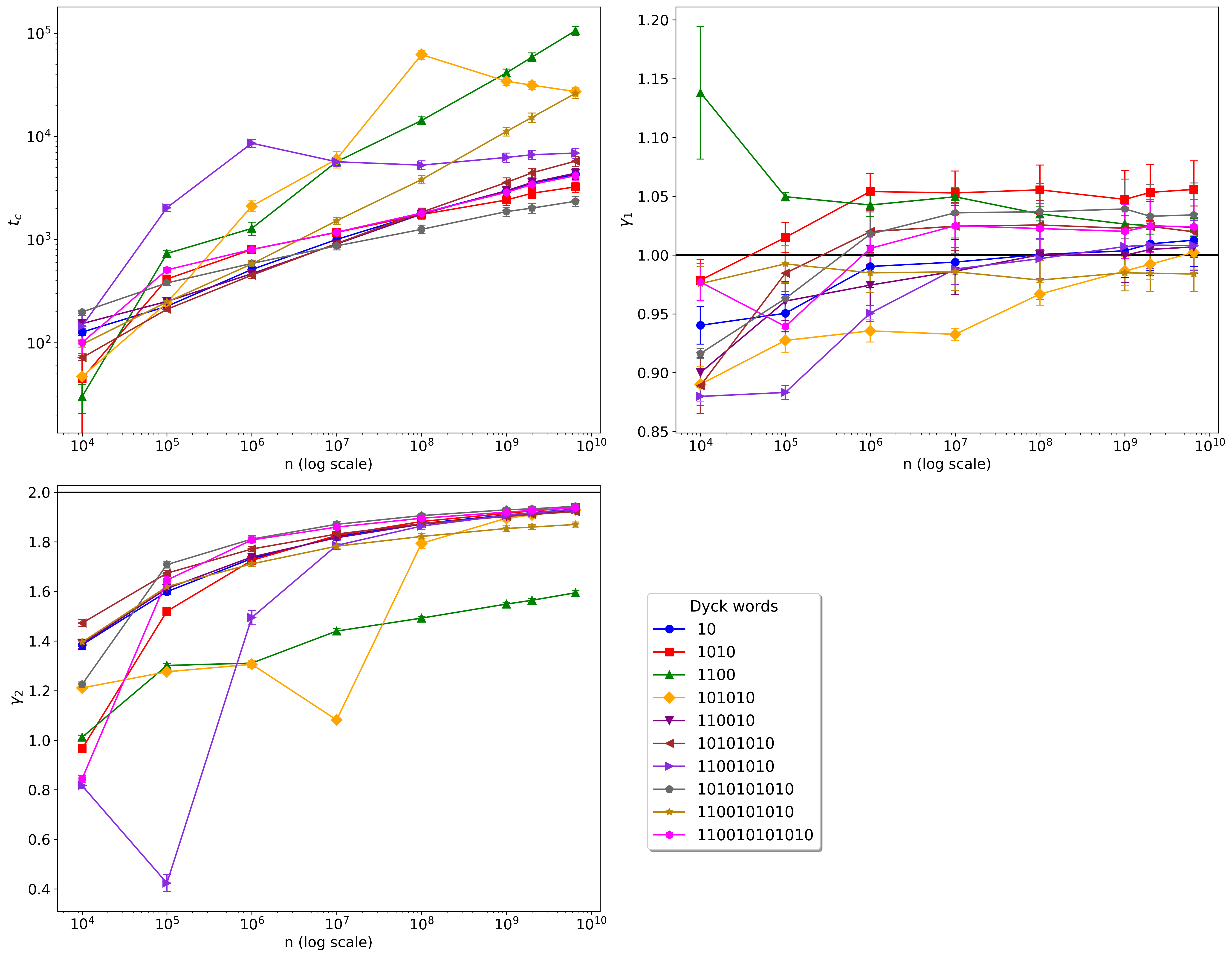}
  \caption{Parameters $t_c$ (left upper panel), $\gamma_1$ (right upper panel) and $\gamma_2$ (left lower panel) for several Dyck words (see the legend) computed  on texts ${\mathbb{N}\mathcal{T}}_{1}^{n}$ of increasing length $n$. The horizontal axis represents the logarithm of $n$.}
  \label{fig:enter-label}
\end{figure}

We have computed the mean square displacement for several Dyck words, obtaining some evidence of the existence of two distinct qualitative behaviours related to the nature of the words themselves. 
\begin{itemize}
\item Let us consider the Dyck words corresponding to {\em square-free numbers}, that is, natural numbers whose factorisation \eqref{eq:factorisation} contains each prime factor no more than once (i.e. $n_1=\cdots=n_k=1$). The trees corresponding to such numbers are ``bushes" and their Dyck words are sequences of \texttt{10} blocks. We consider the following examples: $$\rho= \texttt{10,\,1010,\,101010,\,10101010,\,1010101010}.$$ The 
${\rm MSD}_\rho(t)$ for these Dyck words are reported in Figure \ref{msd_gamma_squarefree}. The log-log plots present piecewise linear behaviours, consistent with those described by equation\eqref{eq:mad-crossover}, with parameters $\gamma_1, \gamma_2, t_c$ depending on the considered Dyck word. In all examples, $\gamma_1$ is slightly greater than 1, but with confidence intervals of the fit that include it, while the exponent $\gamma_2$ is slightly smaller than 2. On the other hand, the crossover point $t_c$ seems quite sensitive to the Dyck word $\rho$. Then, by adhering to the analogy with transport phenomena, we may say that walks associated with square-free numbers have a crossover from approximately normal or slightly superdiffusive to quasi-ballistic behaviours as the lag $t$ crosses some  critical value $t_c$.
\item Some Dyck words associated with {\em non-square-free numbers}, i.e. $$\rho= \texttt{1100, 110010, 11001010, 1100101010, 110010101010},$$ are considered in Figure \ref{msd_gammas_nonsquarefree}. Here too, the piecewise linear behaviour (in log-log plot) is evident, although with some differences compared to the previous case. In particular, while $\gamma_1$ is very close to 1—slightly higher in some cases and slightly lower in others— $\gamma_2$ is consistently less than 2, even when the fit confidence interval is taken into account. 
In these cases, an almost normal behaviour is followed by a superdiffusive one (although not ballistic) as $t$ increases through $t_c$.
%Thus, in light of these examples, one may hypothesize that in the case of non-square-free trees, the crossover point $t_c$
% separates a normal regime from a subdiffusive one.
\end{itemize}
The evaluation of the exponents $\gamma_1$, $\gamma_2$, and of the crossover point $t_c$ depends on the length $n$ of the text ${\mathbb{N}\mathcal{T}}_{1}^{n}$  on which the computation is performed. The results for several values of $n$ are shown in Figure \ref{fig:enter-label}, both for certain Dyck words corresponding to square-free numbers and to non-square-free ones. More precisely, we considered 9 of the 10 most frequent Dyck words in the whole text ${\mathbb{N}\mathcal{T}}_{1}^{\overline{N}}$ (of which 4 are non-square-free), and the Dyck word 1100, which is the shortest non-square-free Dyck word (its  rank is $r<10$ only up to ${\mathbb{N}\mathcal{T}}_{1}^{100}$).

Evidently, a behaviour similar to that described by equation \eqref{eq:mad-crossover} can be observed when considering a text of sufficiently large length $n$. The upper-left panel of Figure \ref{fig:enter-label} shows that the estimated value of the crossover point $t_c$ varies significantly with $n$, and in most cases increases as $n$ grows. While for some Dyck words $t_c$ \invisible{, though increasing,} appears to approach an asymptotic value, in other cases — such as for the Dyck word \texttt{1100} or \texttt{1100101010} — the value of $t_c$ continues to increase. Despite the variability of $t_c$, the values of $\gamma_1$ and $\gamma_2$ appear overall closer to having reached their asymptotic regime, even in cases where $t_c$ is still far from exhibiting limiting behaviour. The upper-right panel in Figure \ref{fig:enter-label} illustrates more explicitly that at short range (i.e. for $t<t_c$), the behaviour varies from one Dyck word to another: in some cases 
$\gamma_1<1$, while in others $\gamma_1>1$, thus revealing a spectrum of dynamics ranging from subdiffusion to weak superdiffusion. At long range ($t>t_c$), see the lower-left panel, the behaviour is clearly superdiffusive, as $\gamma_2$
is significantly greater than 1 — even in the case of the Dyck word \texttt{1100}, where the exponent is markedly smaller than those of the other Dyck words.

Based on the present analysis, we are not in a position to formulate general statements regarding the potential monotonicity properties of the exponents $\gamma_1$ and $\gamma_2$ as functions of $n$. However, we observe that for the most frequent Dyck word (among the large values of $n$ considered), namely $\texttt{101010}$, the exponent $\gamma_2$ exhibits oscillatory behaviour, whereas $\gamma_1$ increases monotonically toward 1. We conclude by noting that $\gamma_1$ displays a variety of behaviours across the different Dyck words considered.

\subsection{Cross-correlation of Dyck words}
An analysis analogous to that presented in Section \ref{sez:MSD} can be carried out by examining the correlation between the walks associated with two distinct Dyck words $\rho_1$ and $\rho_2$. Here we adopt the same notation introduced in Section \ref{sez:MSD}, using the superscript $j$ to denote quantities associated with the Dyck word $\rho_j$. 
%Let ${\bf y}^{(j)},\, j=1,2$ the centered sequences corresponding to $\rho_{j}$ and
Then, given the displacements of the walks ${\bf s}^{(1)}$ and  ${\bf s}^{(2)}$   over a lag $t$:
$$
\Delta s^{(j)}_{n,t} = s^{(j)}_{n+t}-s^{(j)}_n= \sum_{i=n+1}^{n+t} y^{(j)}_i\,,
$$
their {\em cross correlation}  is defined as:
\begin{equation}\label{msd_def SQ}
%\begin{split}
\text{MSD}_{\rho_1,\rho_2}(t) = \left\langle \left( \Delta s^{(1)}_{n,t} - \Delta s^{(2)}_{n,t} \right)^2 \right\rangle 
= \frac{1}{\overline{N} - t} \sum_{n=1}^{\overline{N} - t} \left( \sum_{i=n+1}^{n+t} \left(y^{(1)}_i - y^{(2)}_i\right) \right)^2 . 
%\end{split}
\end{equation}

As for the mean square displacement, in standard situations one may expect the cross-correlation to exhibit a power-law behaviour. However, computations performed for different choices of $\rho_1$ and $\rho_2$ show a double-regime
power-law with a crossover, similar to what is observed in the mean square displacement, see equation\eqref{eq:mad-crossover}.

\begin{figure}[H]
    \centering
    % Row 1
    \begin{subfigure}{0.49\textwidth}
        \centering
        \includegraphics[width=\linewidth]{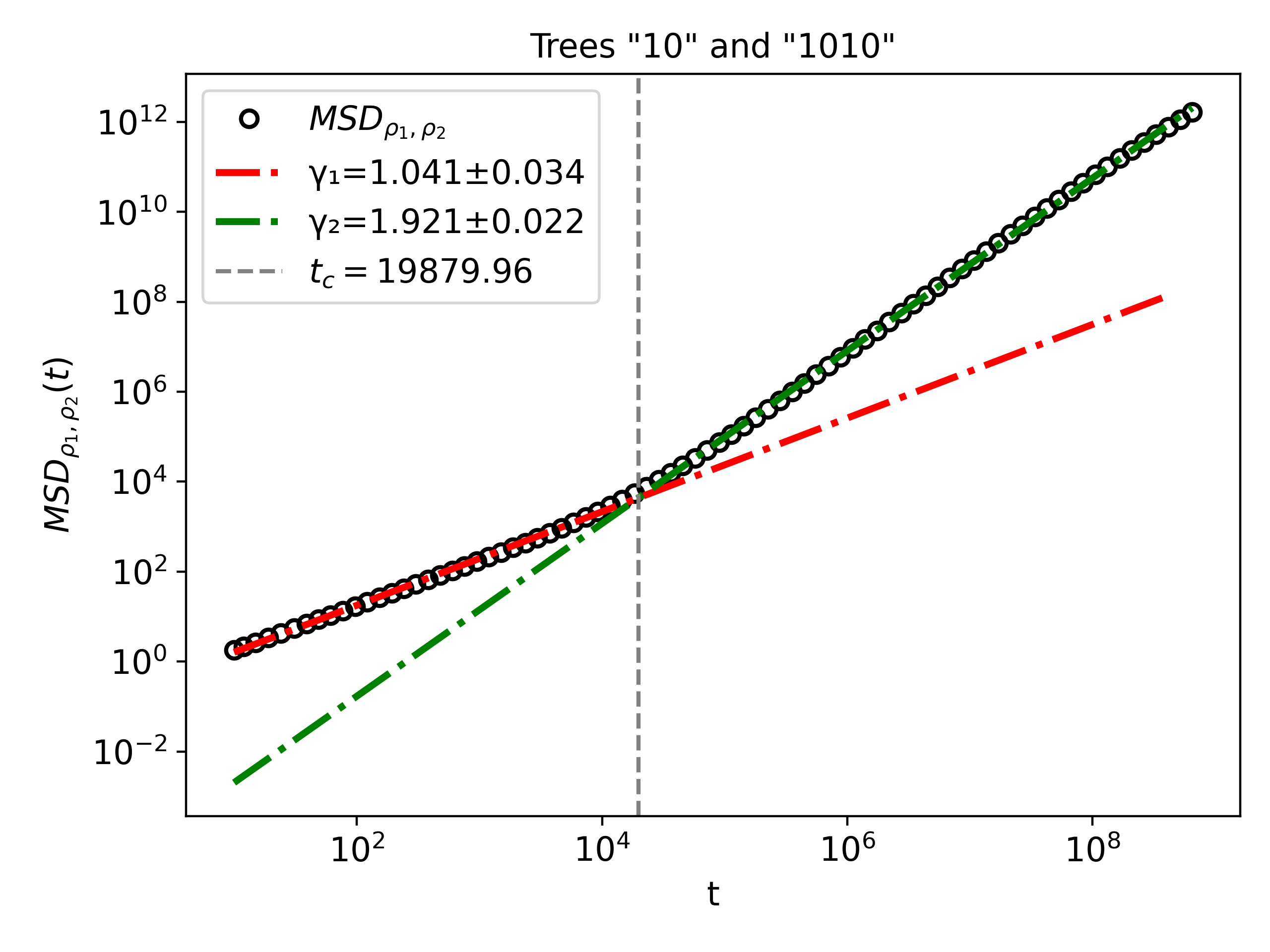}
        \label{MSD_cross1-5}
    \end{subfigure}
    \hfill
    \begin{subfigure}{0.49\textwidth}
        \centering
        \includegraphics[width=\linewidth]{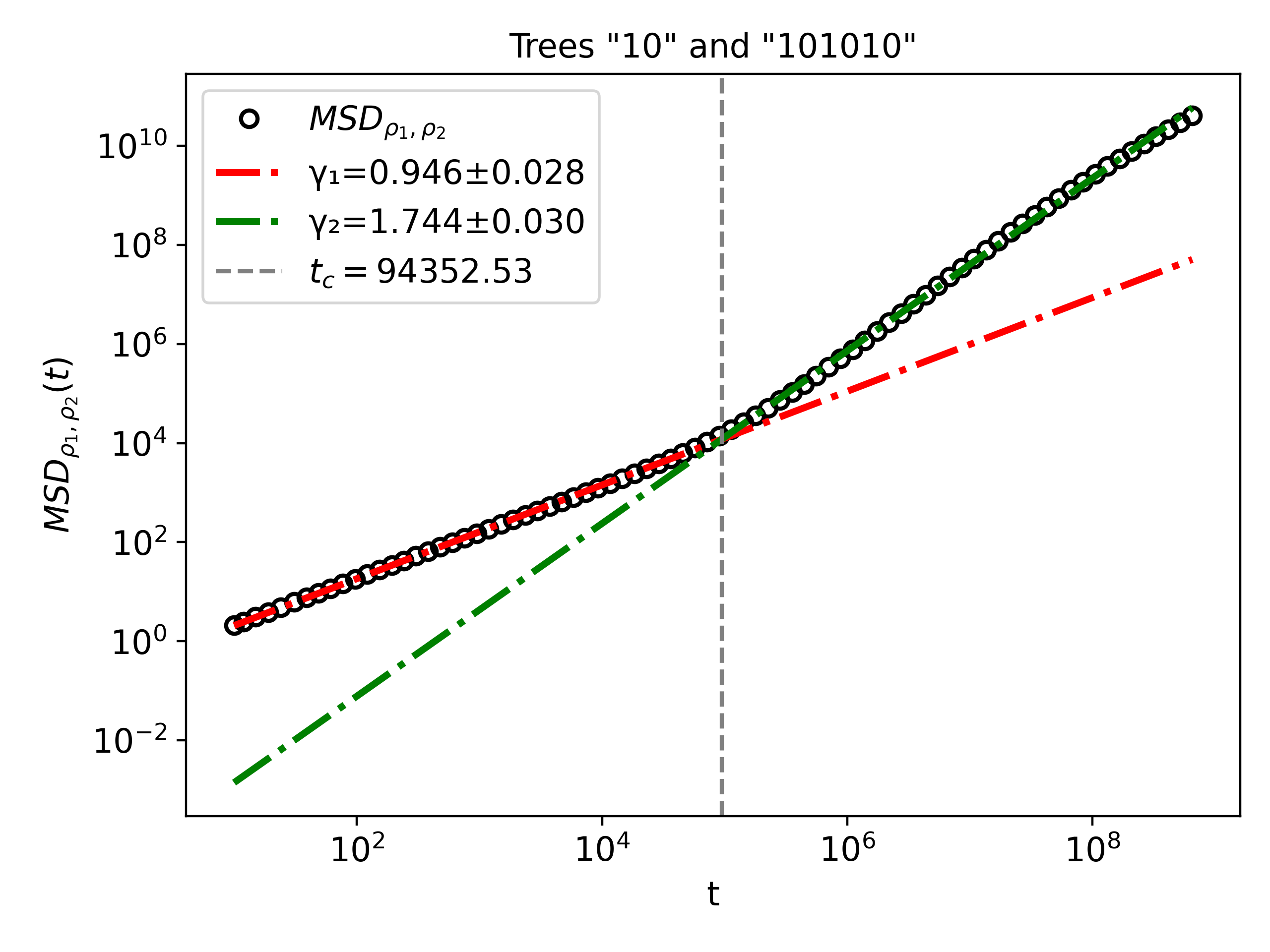}
        \label{MSD_cross1-21}
    \end{subfigure}

    % Row 2
    \begin{subfigure}{0.49\textwidth}
        \centering
        \includegraphics[width=\linewidth]{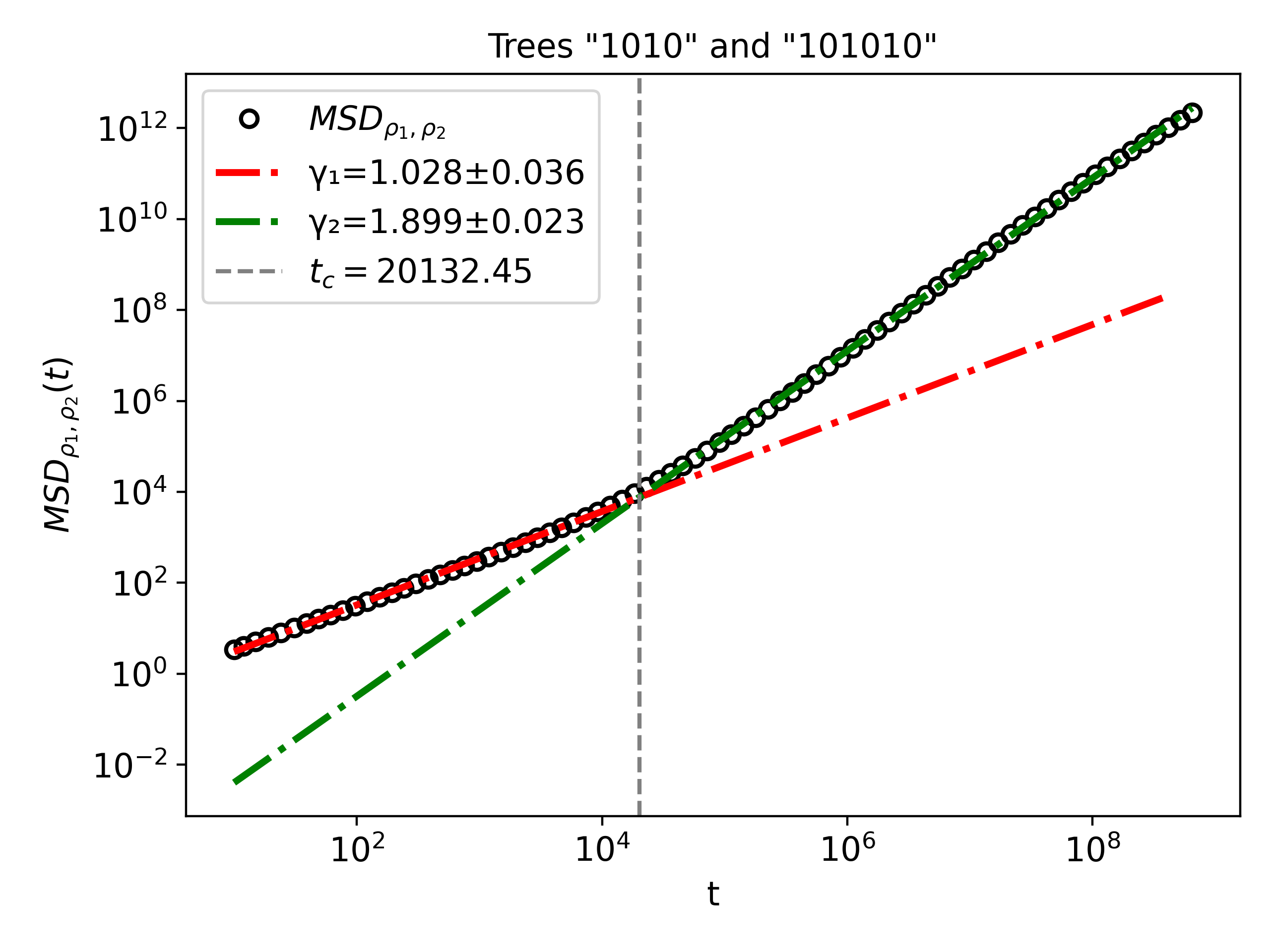}
        \label{MSD_cross5-21}
    \end{subfigure}
    \hfill
    \begin{subfigure}{0.49\textwidth}
        \centering
        \includegraphics[width=\linewidth]{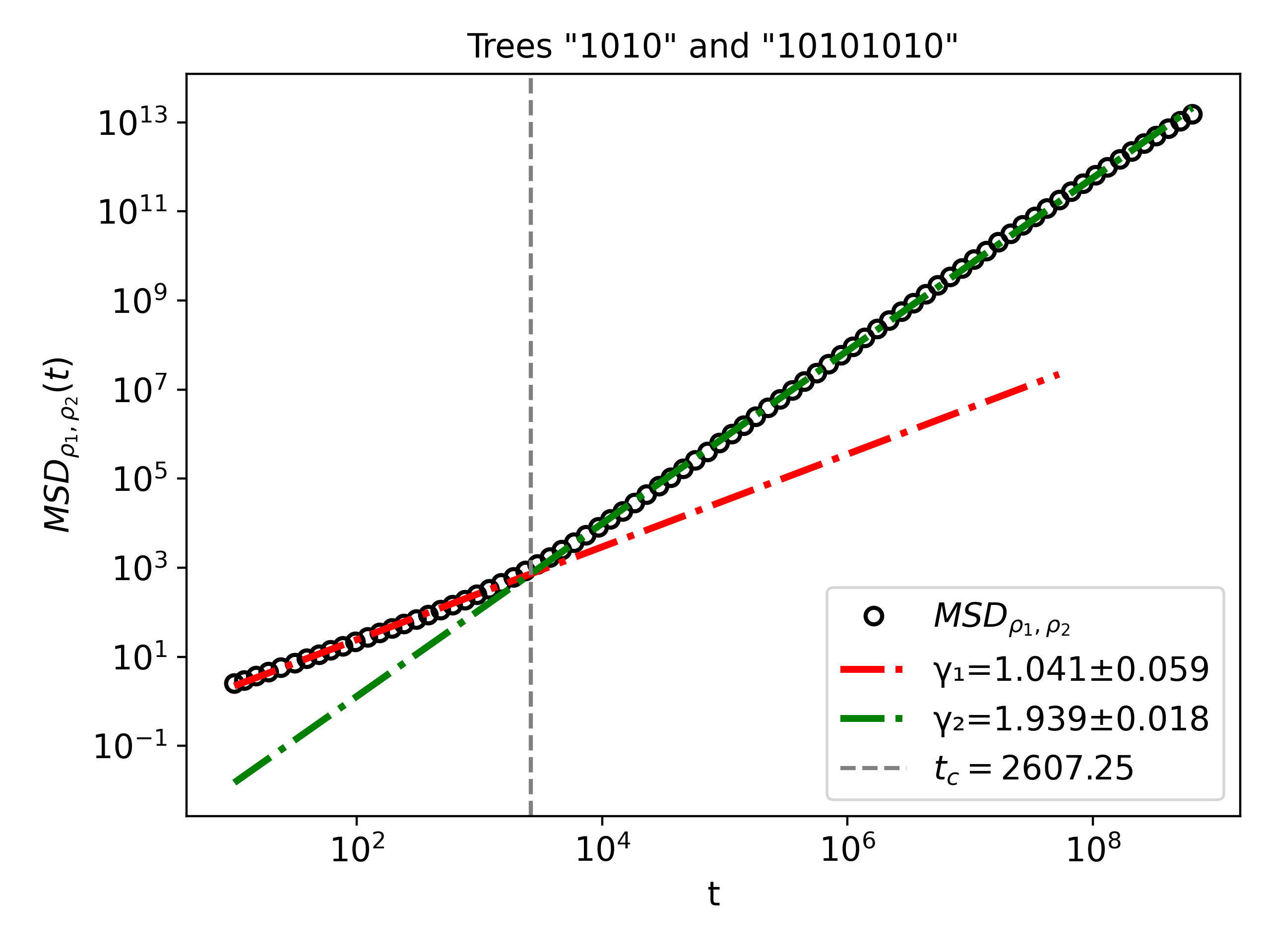}
        \label{MSD_cross5-85}
    \end{subfigure}

    % Row 3
    \begin{subfigure}{0.49\textwidth}
        \centering
        \includegraphics[width=\linewidth]{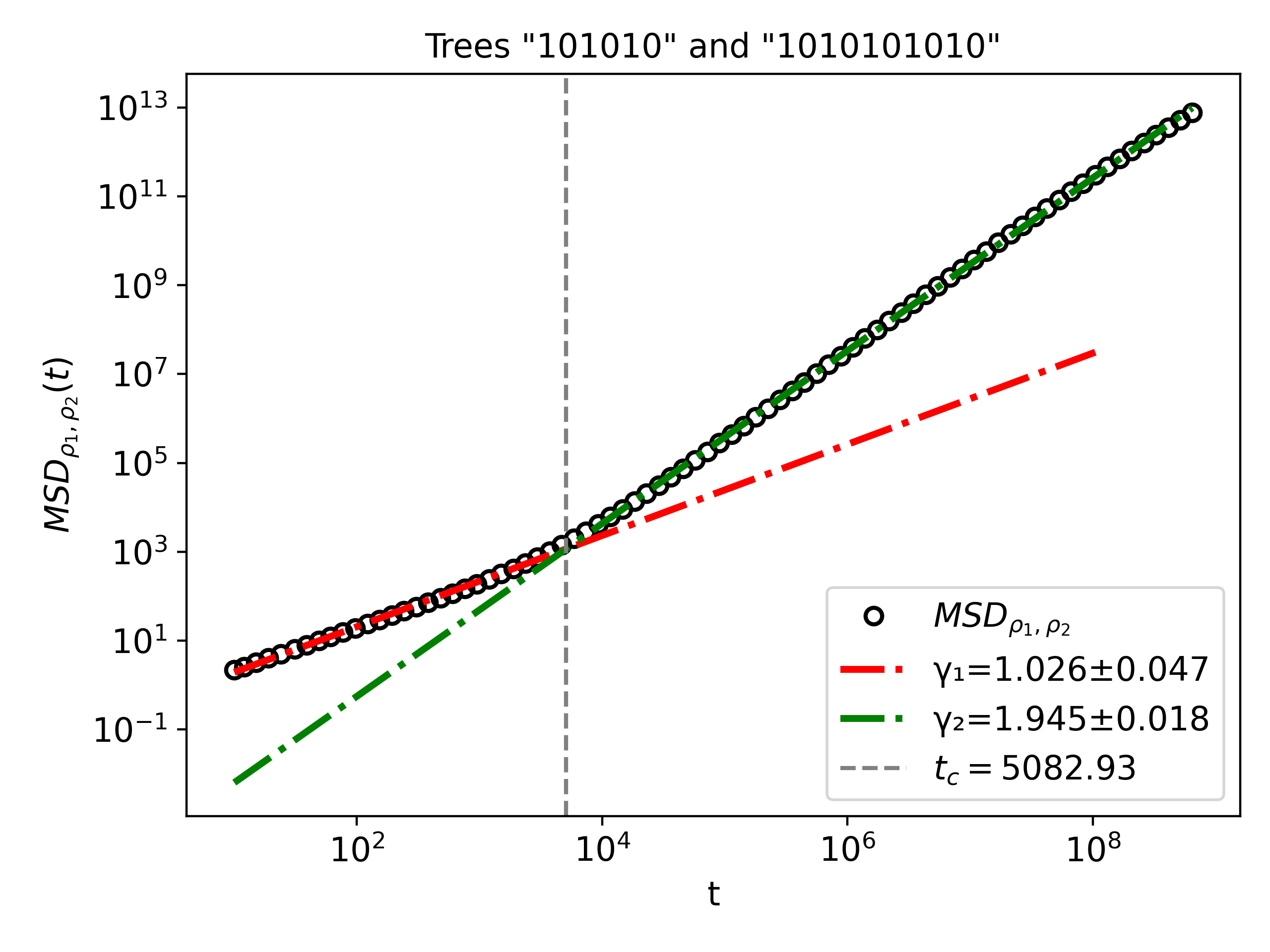}
        \label{MSD_cross21-341}
    \end{subfigure}
    \hfill
    \begin{subfigure}{0.49\textwidth}
        \centering
        \includegraphics[width=\linewidth]{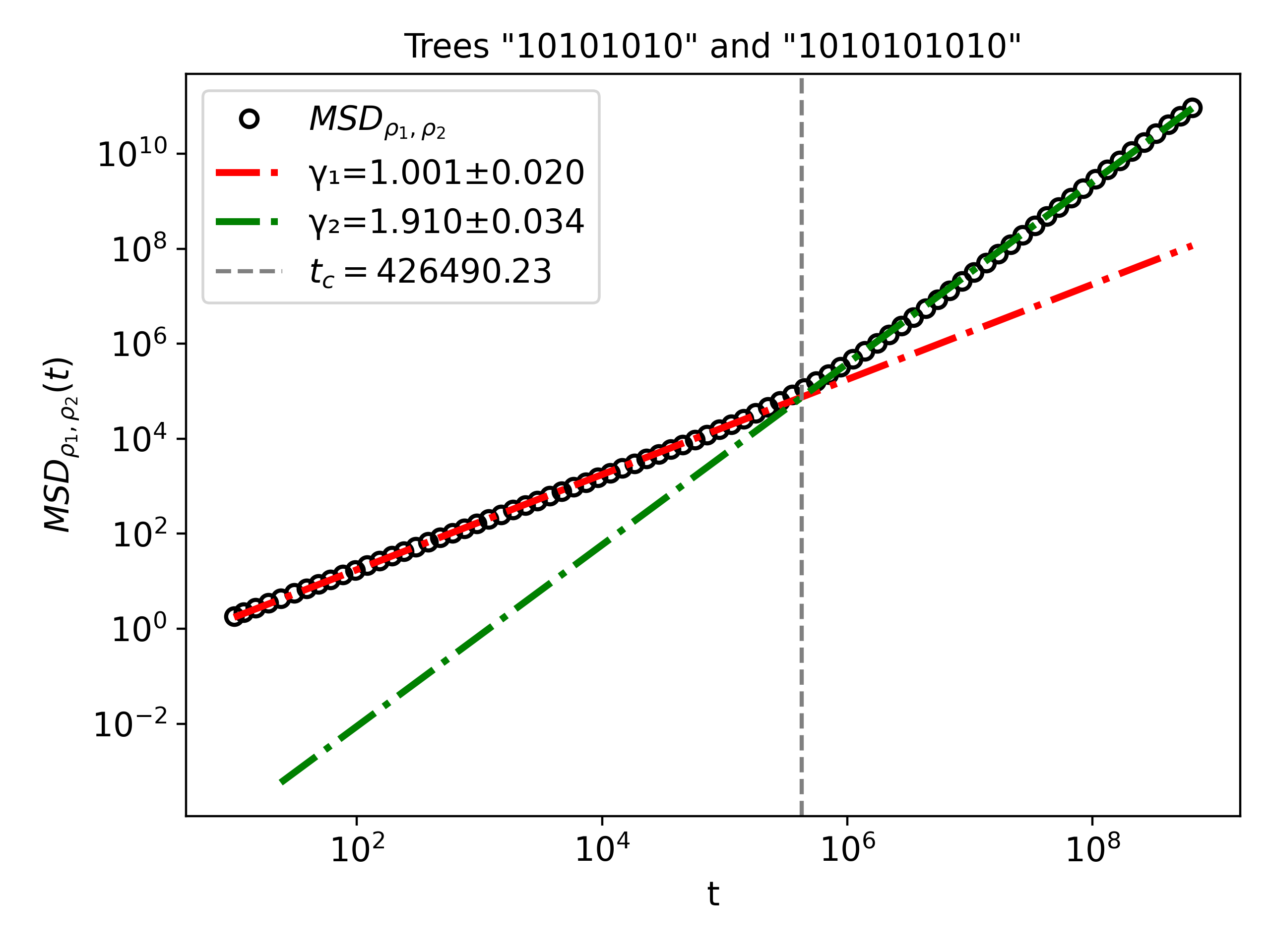}
        \label{MSD_cross85-341}
    \end{subfigure}
    \caption{Cross-correlation $MSD_{\rho_1\rho_2}(t)$ for Dyck words (Trees) $\rho_1$ and $\rho_2$ associated with some square-free numbers versus the time lag $t$ on the log-log scale.}
    \label{cros gamma sqfree}
\end{figure}

\begin{figure}[H]
    \centering

    % Row 1
    \begin{subfigure}{0.49\textwidth}
        \centering
        \includegraphics[width=\linewidth]{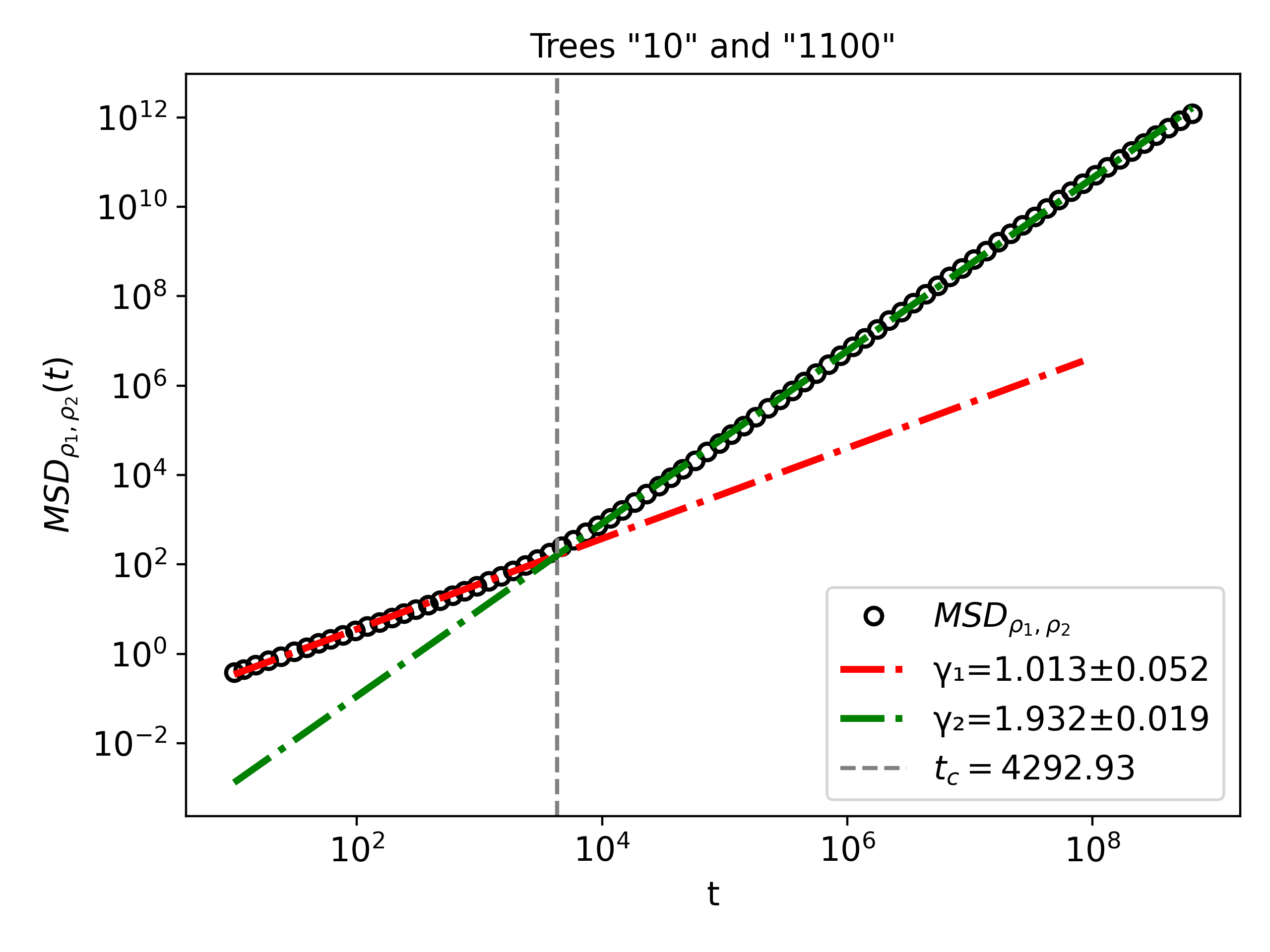}
        \label{MSD_cross1-6}
    \end{subfigure}
    \hfill
    \begin{subfigure}{0.49\textwidth}
        \centering
        \includegraphics[width=\linewidth]{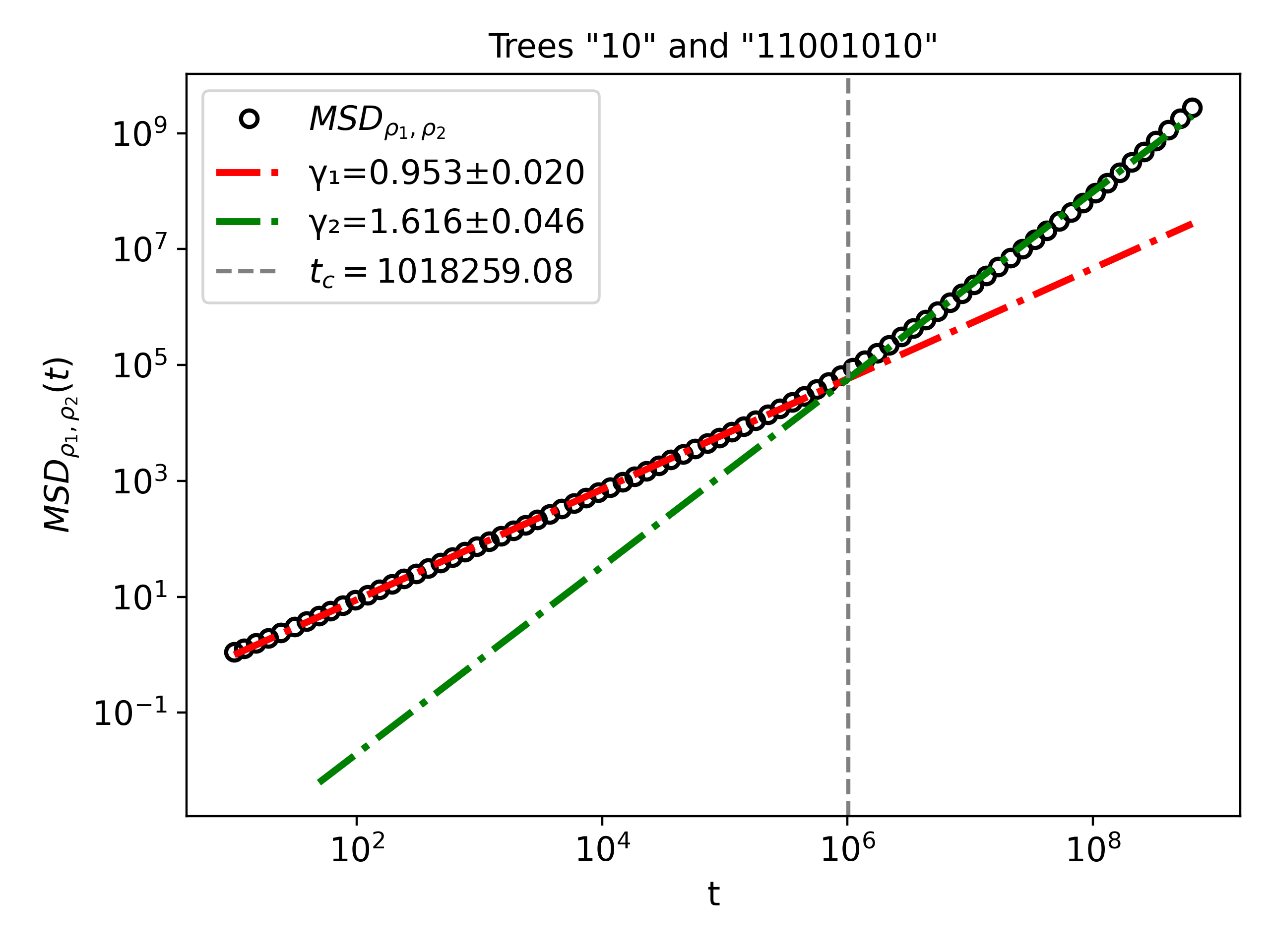}
        \label{MSD_cross1-101}
    \end{subfigure}

    % Row 2
    \begin{subfigure}{0.49\textwidth}
        \centering
        \includegraphics[width=\linewidth]{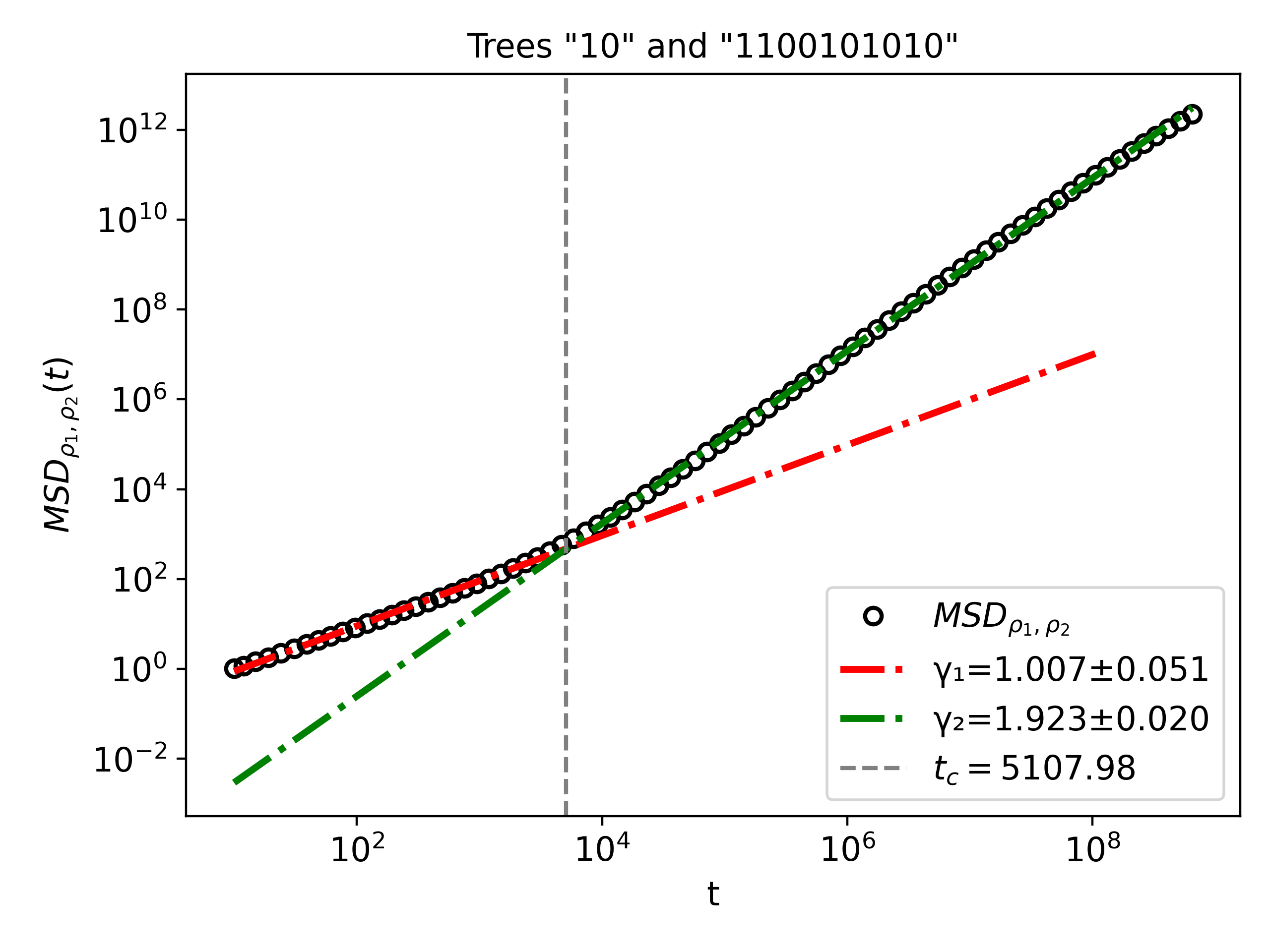}
        \label{MSD_cross1-405}
    \end{subfigure}
    \hfill
    \begin{subfigure}{0.49\textwidth}
        \centering
        \includegraphics[width=\linewidth]{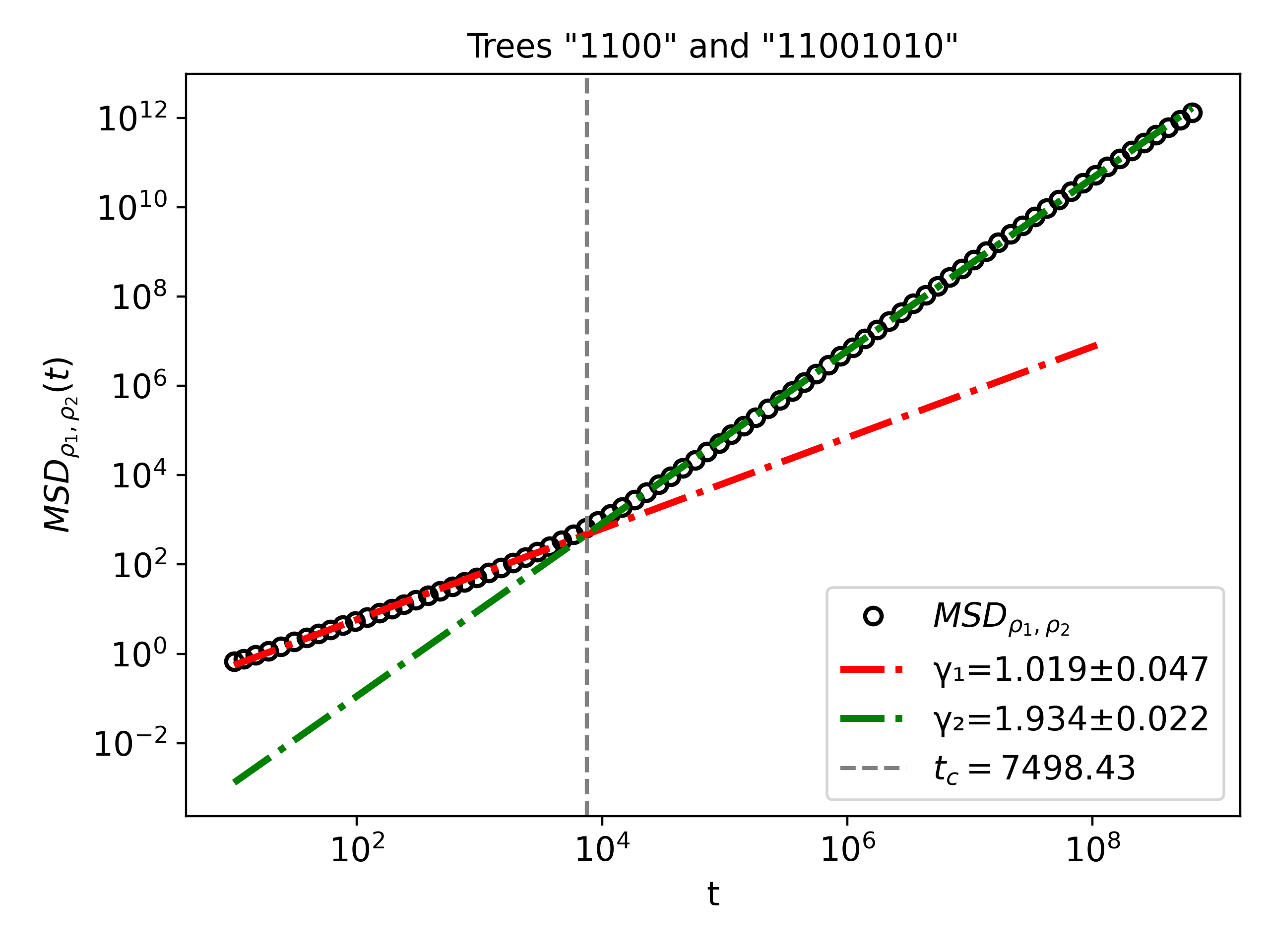}
        \label{MSD_cross6-101}
    \end{subfigure}

    % Row 3
    \begin{subfigure}{0.49\textwidth}
        \centering
        \includegraphics[width=\linewidth]{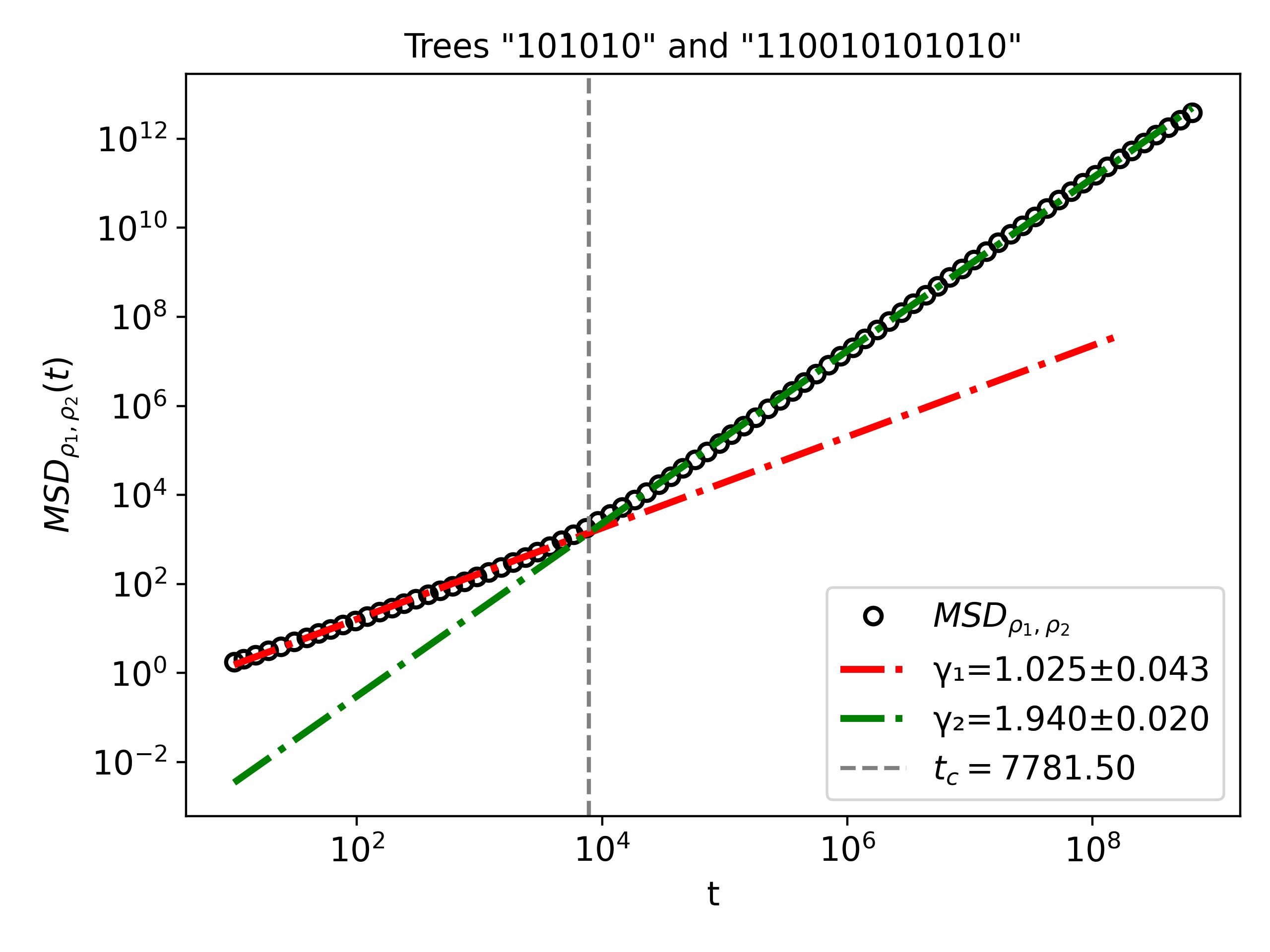}
        \label{MSD_cross21-1621}
    \end{subfigure}
\caption{Cross-correlation $MSD_{\rho_1\rho_2}(t)$ for Dyck words (Trees) $\rho_1$ and $\rho_2$ associated with some non-square-free numbers versus the time lag $t$ on the log-log scale.}
\label{cross gamma nonsq-free}
\end{figure}

Figures \ref{cros gamma sqfree} and 
\ref{cross gamma nonsq-free} show the cross-correlations computed for some of the most frequent Dyck words in \Nd. Specifically, the former considers pairs of square-free numbers, while the latter focuses on pairs consisting of one square-free and one non-square-free number, or two non-square-free numbers. The figures suggest that, at short range, the behaviour is weakly superdiffusive (though subdiffusive in one case), while at long range it becomes superdiffusive.

To provide insight into the  variability of the exponents, we report some of their values in the tables  below (see Table \ref{gammas}), which refer to the five Dyck words corresponding to square-free numbers: $\rho_1=\texttt{10}$, $\rho_2= \texttt{1010}$, $\rho_3=\texttt{101010}$, $\rho_4=\texttt{10101010}$ and $\rho_5=\texttt{1010101010}$. At the intersection of the row $i$ and the column $j$, one finds the exponent $\gamma$ of $MSD_{\rho_i,\rho_j}(t)$.

\begin{table}[H]
\centering
\caption{Scaling exponents $\gamma$ for square-free Dyck words.}
\begin{tabular}{|c|ccccc|}
\hline
 & \multicolumn{5}{c|}{\textbf{Short-time scaling ($\gamma_1$)}} \\
\cline{2-6}
\textbf{Dyck Word} & \texttt{10} & \texttt{1010} & \texttt{101010} & \texttt{10101010} & \texttt{1010101010} \\
\hline
\texttt{10} & 1.013 & 1.041 & 0.946 & 1.007 & 1.049 \\
\texttt{1010} & 1.041 & 1.056 & 1.028 & 1.041 & 1.053 \\
\texttt{101010} & 0.946 & 1.028 & 1.005 & 1.049 & 1.026 \\
\texttt{10101010} & 1.007 & 1.041 & 1.049 & 1.030 & 1.001 \\
\texttt{1010101010} & 1.049 & 1.053 & 1.026 & 1.001 & 1.049 \\
\hline
\end{tabular}

%\vspace{0.5cm}

\begin{tabular}{|c|ccccc|}
\hline
 & \multicolumn{5}{c|}{\textbf{Long-time scaling ($\gamma_2$)}} \\
\cline{2-6}
\textbf{Dyck Word} & \texttt{10} & \texttt{1010} & \texttt{101010} & \texttt{10101010} & \texttt{1010101010} \\
\hline
\texttt{10} & 1.932 & 1.921 & 1.744 & 1.934 & 1.946 \\
\texttt{1010} & 1.921 & 1.939 & 1.899 & 1.939 & 1.949 \\
\texttt{101010} & 1.744 & 1.899 & 1.933 & 1.929 & 1.945 \\
\texttt{10101010} & 1.934 & 1.939 & 1.929 & 1.925 & 1.910 \\
\texttt{1010101010} & 1.946 & 1.949 & 1.945 & 1.910 & 1.946 \\
\hline
\end{tabular}\label{gammas}
\end{table}

%\newpage

\section{Conclusion and Outlook}\label{sec:conclusion}

The analysis presented in this work examines the arithmetic sequence through its planar rooted-tree representation and, equivalently, through the associated symbolic text of Dyck words. The text is generated by iterating the Euclidean decomposition of natural numbers down to a prime-only representation, and is therefore a fully deterministic corpus. Its composition naturally includes primes and all other trees, and the resulting distribution displays structural organization across multiple scales. Remarkably, these regularities emerge without any stochastic assumptions or imposed generative models, revealing intrinsic features of the sequence itself.
We found that the dictionary grows sublinearly, indicating sustained reuse of a limited set of combinatorial patterns. The entropy increases slowly and remains well below the theoretical bound, reflecting a high degree of redundancy. The compression ratio shows a non-monotonic behaviour, signalling a transition from local regularity to a broader form of organized complexity. The rank–frequency distribution is stable in shape and departs markedly from a Zipf law, being well approximated instead by a parabolic profile, in log-log plot, over several orders of magnitude in rank. 
The observed curvature, consistent with self-similarity between ranks \cite{Sornette1997, Montemurro2001}, is analogous to departures from the classical Zipf behaviour documented in natural languages \cite{Liu2017, Deng2013, Liu_Nuo_Wu_2014, Ha_Hanna_Ming_Smith_2009}.
 
Transitions of the MSD from short–lag diffusion to enhanced scaling have been reported in
correlated systems. Analogous behaviours were observed in dense active Brownian suspensions
\cite{Reichert2021} and in disordered porous media \cite{Li2006, Gmachowski2015}. A similar
phenomenology was also identified in symbolic sequences, where long–range correlations in texts yield superdiffusive MSD exponents \cite{AltCriDesp}. These
examples indicate that MSD scaling beyond normal diffusion can arise in structured or
correlated settings.

A different type of crossover occurs in processes where the transport exponent decreases
with scale. In models with memory kernels, \cite{IlyinProcaccia2010} reported
a transition from ballistic to fractional–diffusive motion; a similar direction was found in
active turbulence \cite{SinghChaudhuri2024}, and in ageing CTRW models where long–time
behaviour becomes subdiffusive \cite{Sokolov2012}. Here, the analogy lies not only in the presence of scale-dependent exponents but also in the fact that the short-time behavior is approximately normal in both cases, rather than in the asymptotic transport regime.

Several directions follow naturally.
First, the empirical regularities observed here call for theoretical explanations, particularly concerning the origin of the parabolic rank–frequency law and the mechanisms that determine the two-regime correlation structure. Second, the planar rooted-tree text offers a fully controlled benchmark for studying learnability in transformer models \cite{BolognaLLM2025}, where the role of determinism, hierarchy, and redundancy can be isolated. Finally, connecting the combinatorial depth of natural numbers with linguistic-type observables suggests a broader programme: to understand how the arithmetic structure expresses itself when viewed as a symbolic language.
This study provides the empirical baseline for such developments.

\section*{Acknowledgements}
The authors are grateful to Gabriele Sicuro for providing the tree-encoded database of the natural numbers. This research was performed under the auspices of Italian National Group of Mathematical
Physics (GNFM) of the National Institute for Advanced Mathematics - INdAM.
The authors acknowledge the financial support from the European Union - Next Generation EU - Grant PRIN 2022B5LF52. This project received support from the EU H2020 ICT48 project Humane AI Net (grant no. 952026), the Italian Ministry of University and Research PRIN 2022 (code J53D23003690006), and the Italian Extended Partnership PE01—FAIR (Future Artificial Intelligence Research, proposal code PE00000013) under the National Recovery and Resilience Plan. Claudio Giberti is a member of the Interdepartmental Centers En\&Tech and InterMech at the University of Modena and Reggio Emilia, Italy.

%\appendix
%\section{Example Appendix Section}
%\label{app1}


%%       \cite{<label>} ==> [1]

%%

\bibliographystyle{elsarticle-num} 
\bibliography{ref}



\end{document}